\documentclass[sigconf]{acmart}

\newcommand{\ignore}[1]{}
\usepackage{enumitem}
\usepackage{multirow}
\usepackage{physics}
\usepackage{amsmath}
\usepackage[noend,linesnumbered,ruled,vlined]{algorithm2e}

\AtBeginDocument{%
  \providecommand\BibTeX{{%
    \normalfont B\kern-0.5em{\scshape i\kern-0.25em b}\kern-0.8em\TeX}}}

\setcopyright{acmcopyright}
\copyrightyear{2021} 
\acmYear{2021} 
\setcopyright{acmcopyright}\acmConference[MICRO '21]{MICRO'21: 54th Annual IEEE/ACM International Symposium on Microarchitecture}{October 18--22, 2021}{Virtual Event, Greece}
\acmBooktitle{MICRO'21: 54th Annual IEEE/ACM International Symposium on Microarchitecture (MICRO '21), October 18--22, 2021, Virtual Event, Greece}
\acmPrice{15.00}
\acmDOI{10.1145/3466752.3480044}
\acmISBN{978-1-4503-8557-2/21/10}

\begin{document}

\title{JigSaw: Boosting Fidelity of NISQ Programs\\via Measurement Subsetting}

\author{Poulami Das}
\email{poulami@gatech.edu}
\affiliation{%
  \institution{Georgia Institute of Technology}
  \city{Atlanta}
  \country{USA}
}

\author{Swamit Tannu}
\email{stannu@wisc.edu}
\affiliation{%
  \institution{University of Wisconsin}
  \city{Madison}
  \country{USA}
  }

\author{Moinuddin Qureshi}
\email{moin@gatech.edu}
\affiliation{%
  \institution{Georgia Institute of Technology}
  \city{Atlanta}
  \country{USA}
}

\renewcommand{\shortauthors}{Das, Tannu, and Qureshi}

\begin{abstract}
Near-term quantum computers contain noisy devices, which makes it difficult to infer the correct answer even if a program is run for thousands of trials. On current machines, qubit measurements tend to be the most error-prone operations (with an average error-rate of 4\%) and often limit the size of quantum programs that can be run reliably on these systems. As quantum programs create and manipulate correlated states, all  the program qubits are measured in each trial and thus, the severity of measurement errors increases with the program size. The fidelity of quantum programs can be improved by reducing the number of measurement operations.

We present {\em JigSaw}, a framework that reduces the impact of measurement errors by running a program in two modes. First, running the entire program and measuring all the qubits for half of the trials to produce a global (albeit noisy) histogram.  Second, running additional copies of the program and measuring only a subset of qubits in each copy, for the remaining trials, to produce localized (higher fidelity) histograms over the measured qubits. JigSaw then employs a Bayesian post-processing step, whereby the histograms produced by the subset measurements are used to update the global histogram.  Our evaluations using three different IBM quantum computers with 27 and 65 qubits show that JigSaw improves the success rate on average by 3.6x and up-to 8.4x. Our analysis shows that the storage and time complexity of JigSaw scales linearly with the number of qubits and trials, making JigSaw applicable to programs with hundreds of qubits.

\end{abstract}




\maketitle

\section{Introduction}
Quantum computers can solve very hard problems by using properties of quantum bits (qubits)~\cite{lloyd1996universal,shor1999polynomial}. Recently demonstrated quantum hardware with fifty-plus qubits are getting to the regime where they can outperform the most advanced supercomputer for some computations~\cite{QCSup}. Unfortunately, these machines are not sufficiently large to implement quantum error correction and are operated in the {\em Noisy Intermediate Scale Quantum (NISQ)}~\cite{preskillNISQ} model, whereby the computation is run a large number of times (called trials) and the answer is inferred from the output log. The ability to obtain the correct answer on a NISQ machine depends on the error-rates and the size of the program. Recently, various software techniques have been investigated to improve application fidelity that either perform noise-aware computations to enable better than worst-case error rate~\cite{noiseadaptive,murali2020software,micro1,FNM,tannu2019not} or reduce the program length and number of computations~\cite{li2018tackling,shi2019optimized,zulehner2018efficient}.

This paper focuses on measurement operations, which are often the most dominant source of error on current superconducting quantum computers, with average error-rates ranging between 2-7\%~\cite{ibmqsystems,QCSup}. Measurement errors constrain the size of the largest program that can be run on NISQ machines with high fidelity~\cite{geller2020towards}. These errors arise from the imperfections in the qubit readout protocol~\cite{cqedschoelkopf,cqed} and the long latency of these operations (about 4-5 microseconds on IBM quantum systems). Furthermore, measurement operations suffer from {\em measurement crosstalk}~\cite{arute2019supplementary,khezri2015qubit}, which means performing several measurement operations concurrently increases the error rate of each measurement operation. Our experiments on IBMQ machines show that the average measurement error rate increases by up-to 2\% when five qubits  (and by up-to 4\% when ten qubits) are measured simultaneously compared to measuring a single qubit in isolation. This indicates larger programs become even more susceptible to measurement crosstalk due to higher number of measurement operations. Similar observations are reported for the Google Sycamore, where simultaneous measurement operations have 1.26x higher error-rates compared to isolated measurements~\cite{QCSup}. Consequently, fast and accurate qubit measurement at scale remain an open problem~\cite{krantz2019quantum}.

\ignore{Measurement operations are \textbf{(1)} vulnerable to crosstalk and \textbf{(2)} exhibit high spatial variability in the error-rates. Measurement crosstalk depends on qubits and the quantum state deteriorates the fidelity of these operations when several qubits are measured simultaneously. To understand the impact of measurement crosstalk, we execute the two circuits shown in Figure~\ref{fig:ibmqtoronto}, measuring one and two qubits respectively, on 27-qubit IBMQ-Toronto. The circuits prepare arbitrary quantum states using the single-qubit U3 gates and specifying the Euler angles~\cite{abhijith2018quantum}. We execute the circuits shown in Figure~\ref{fig:ibmqtoronto}(a) and (b) on each qubit and each pair of qubits of the hardware respectively. We compute the \textit{Total Variation Distance (TVD)}~\cite{TVD} between the output states obtained on the noisy hardware and the output on a noise-free quantum computer for each qubit. The TVD is often used to measure the fidelity of quantum states and a lower TVD is better (0 corresponds to no error)~\cite{sanders2015bounding}. We observe that measurement crosstalk increases the TVD by up-to 50x when two qubits are measured simultaneously as opposed to isolated measurements (diagonals in Figure~\ref{fig:ibmqtoronto}(c) and (d)). Furthermore, the impact of crosstalk depends on the quantum state and cannot be characterized a-priori. 

\begin{figure}[htb]
\centering

    \includegraphics[width=\columnwidth]{./Figures/tvd_heatmap.pdf}
    \vspace{-0.25in}
    \caption{Circuit with (a) single measurement (b) two measurements. (c-d) Relative Total Variational Distance of the output for two different quantum states.} 
     \vspace{-0.10 in}
    \label{fig:ibmqtoronto}
\end{figure}
}
\begin{figure*}[htb]
\vspace{-0.15in}
\centering
  \includegraphics[width=\linewidth ]{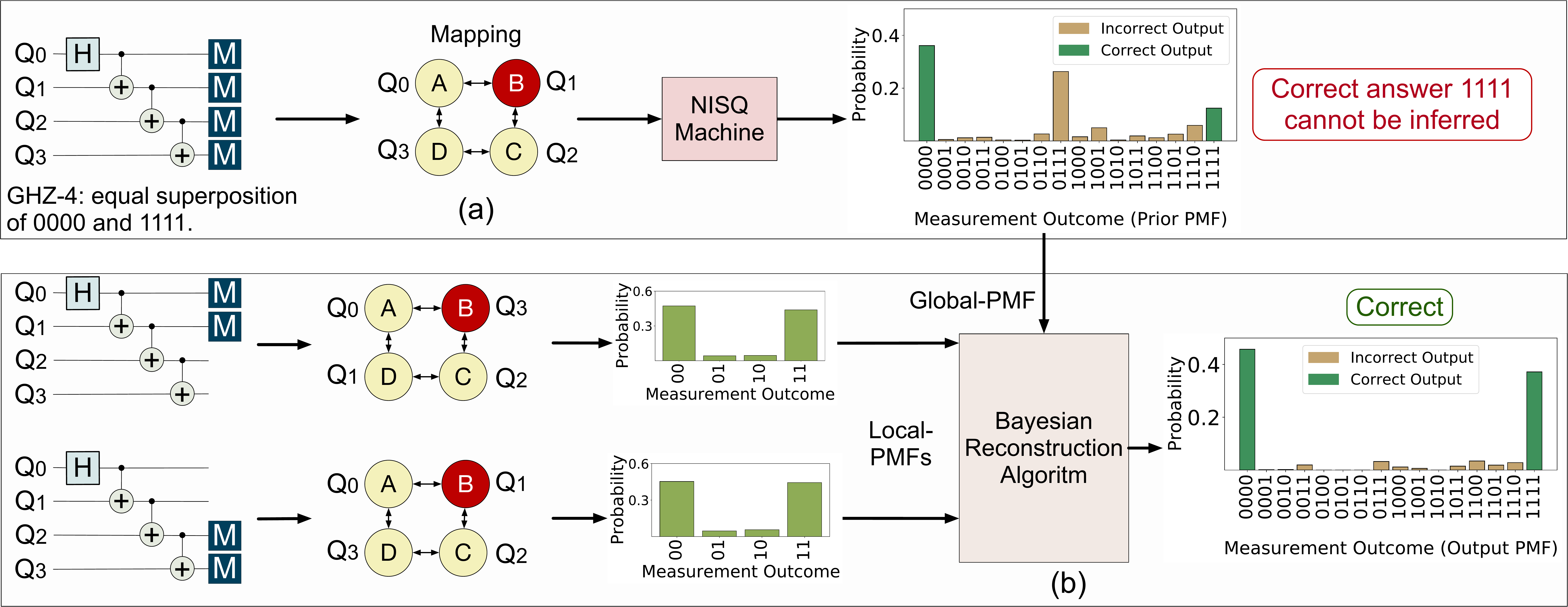}
\caption{(a) Executing GHZ-4 kernel on NISQ machine (b) JigSaw obtains a global-PMF for the program by measuring all the qubits, a high-fidelity local-PMF for each circuit with partial measurement (CPM), and uses the local-PMFs from CPMs to update the global-PMF via a Bayesian Reconstruction algorithm.}
\label{fig:intro}
\vspace{-0.15in}
\end{figure*}

Spatial variation of measurement errors imposes further challenges in attaining high fidelity for large programs. For example, on IBM's 27-qubit Toronto device, the median measurement error rate is about 2.7\%, whereas the worst-case error rate is 22.2\%. Existing state-of-the-art noise-aware compilers avoid mapping programs onto unreliable physical qubits~\cite{noiseadaptive,tannu2019not}, to alleviate the impact of worst-case errors. However, it is not always feasible for large programs, particularly in the context of measurement errors because the qubits with the lowest measurement error rates are typically not co-located in space. For example, it is challenging to map any program with more than six qubits on IBMQ-Toronto without using a physical qubit with more than 2.7\% measurement error-rate (the median). In the limiting case, the compiler is forced to use qubits with more than 20\% measurement error-rate for any program that uses sixteen qubits or more.

NISQ programs measure all the program qubits in each trial, and thus, suffer from the accumulated measurement errors across all the program qubits. The severity of measurement errors is compounded in large programs by the fact that a correct output is obtained only if all these measurements are error-free.  The key insight in our paper is to measure only a subset of program qubits for some of the trials, so that these trials encounter fewer errors due to reduced measurement crosstalk, and by avoiding the physical qubits most susceptible to measurement errors. However, subset measurements lack information about the global correlation across all the qubits. Thus, we need a way to combine the benefits of both global runs (full correlation but low fidelity) and subset runs (limited correlation over fewer qubits but higher fidelity).  To that end, we propose JigSaw,\footnote{The name is inspired by the popular Jigsaw puzzle -- the global-PMF represents a noisy skeleton, and the subset measurements are small tiles. When both are used together, the full picture is revealed.} a framework that reduces measurement errors by running a program in two modes: global and subset. The PMF (probability mass function) produced for each mode are then combined using a post-processing algorithm to generate a higher fidelity PMF.

To illustrate our design, we use the GHZ-4 program shown in Figure~\ref{fig:intro}(a) as an example. Ideally, GHZ-4 produces \texttt{0000} or \texttt{1111} with 50\% probability each.  However, errors can produce incorrect outputs, some with probability higher than the two correct answers. To improve the application fidelity using JigSaw, first, we run the program in {\em global-mode}, wherein we measure all the qubits to produce the global-PMF. In the second step, we run the circuit in {\em subset-mode}, where multiple copies of the original program are run, but where each copy only measures a subset of qubits and produces a PMF only over the measured qubits. For example, in Figure~\ref{fig:intro}(b), we run two such {\em Circuits with Partial Measurements (CPM)} that perform all the computations but only measure two qubits in each copy ($ {\textrm{Q}}_0$ and ${\textrm{Q}}_1$ for the first copy, and $ {\textrm{Q}}_2$ and ${\textrm{Q}}_3$ for the second copy). Each CPM is recompiled to ensure that the two measurement operations are performed on the physical qubits with the least amount of measurement error. Therefore, we would expect the measurement fidelity which is closer to the best-case qubits rather than average-case or worst-case qubits. Measurement subsetting and recompilation results in the higher reliability of CPM marginals compared to the scenario of deriving the marginals from the global-PMF.

Although fewer measurements can offer marginal information with higher fidelity, the CPM may still not be able to produce the output PMF as the correlation between different CPM are unknown. To reconstruct the global-PMF using the local-PMFs, we employ a \textit{Bayesian Reconstruction} algorithm, analogous to Bayesian updating in statistics, where a prior probability estimate is updated using newer information. For JigSaw, the global-PMF is the prior estimate, which is updated using more accurate marginal information from the CPM. By performing Bayesian updates, the algorithm accentuates the probabilities of the correct outcome(s) while reducing the probabilities of the incorrect outcomes, thus improving the fidelity. Our evaluations with three different quantum hardware from IBM, with 27 and 65 qubits, show that JigSaw without recompilation (and measurement subsetting only) improves the success rate of typical quantum benchmarks on average by 1.85x and by up-to 3.26x. Note that by default JigSaw uses CPM of size 2. Alternately, JigSaw with both measurement subsetting and CPM recompilation improves the fidelity of applications on average by 2.9x and by up-to 7.9x.

For an N-qubit program, it is possible to design a polynomial number of CPM of subset size 2 ($^\textrm{N}C_2$). However, our experiments show that 
the effectiveness of JigSaw saturates even when more CPM of the same subset size are used because after a certain limit these additional CPM do not offer incremental and unique information. To overcome this limitation, we propose {\em Multi-Layer JigSaw  (JigSaw-M)}, which creates CPM of different sizes by exploiting the trade-off between fidelity and the amount of correlation for a CPM. While small CPM provide higher fidelity; they do not capture sufficient correlation. Alternatively, larger CPM provide greater correlation but also encounters more errors due to larger number of measurements. Overall, our default JigSaw-M uses CPM of subset sizes 2 to 5 qubits and improves application success rate on average by 3.65x and by up-to 8.42x.

\vspace{0.1 in}

Overall, our paper makes the following contributions:


\begin{enumerate}[leftmargin=0cm,itemindent=.5cm,labelwidth=\itemindent,labelsep=0cm,align=left, itemsep=0.15cm, listparindent=0.3cm]

\item We propose {\em JigSaw}, a framework that does not subject all trials to measurement errors across all qubits. It performs half of the trials with global measurements (for correlation) and the other half with subset measurements (for higher fidelity).

\item We propose a {\em Bayesian Reconstruction} algorithm that uses the PMFs from the subsets to improve the global-PMF.

\item We propose {\em JigSaw-M}, which generates more unique non-uniform sized CPM to further enhance the global-PMF.
\end{enumerate}

\vspace{0.05 in}

Our scalability analysis of JigSaw show that the storage and time complexity is linear with the number of trials and qubits, making JigSaw applicable to programs with hundreds of qubits. The algorithm for JigSaw and datasets used for evaluations in this paper is available at this \href{https://github.com/pdas36/JigSaw}{\underline{link}}. 

\section{Background}
\subsection{Basics of Quantum Computing}

A qubit is the fundamental unit of information on a quantum computer and may be represented using a vector $\ket{\Psi} = \alpha\ket{0} + \beta\ket{1}$, a superposition of the basis states $\ket{0}$ and $\ket{1}$  such that $\mid\alpha\mid^2 + \mid\beta\mid^2 = 1$. When measured, a qubit collapses to a binary $0$ or $1$ with probabilities $\mid\alpha\mid^2$ and $\mid\beta\mid^2$ respectively. Similarly, an  $n$-qubit system exists in a superposition of $2^n$ basis states and can produce any of the $2^n$ bitstrings depending on the probabilities associated with them upon measurement.

\subsection{Errors and NISQ Model of Computing}
Qubits are extremely sensitive devices and error prone. These errors can corrupt their states, producing incorrect outcomes during program execution. The state of a qubit naturally decays due to its interactions with the environment, a phenomenon called decoherence, whereas imperfections in quantum operations can lead to gate errors. \textit{Measurement} errors manifest due to errors in measurement operations. 
Noisy Intermediate Scale Quantum (NISQ) computers will be operated in the presence of noise as they may not be large enough to achieve fault-tolerance~\cite{preskillNISQ}. By repeatedly executing a program several times on the NISQ hardware (called \textit{trials}), the program solution can be inferred from the output distribution.

\subsection{Measurement Errors}
\label{sec:measurementerrorsissues}
Qubit measurement error rates can constrain the size of the largest program (in terms of number of qubits) that can be run reliably on a NISQ machine~\cite{geller2020towards}. As a result, software policies particularly aimed at reducing the impact of measurement errors are currently being developed~\cite{matrixmeasurementmitigation,bravyi2020mitigating,kwon2020hybrid,FNM,barron2020measurement}.  

\newpage
\noindent{\textbf{How are measurements performed?}} Superconducting qubits, similar to the devices from IBM and Google, are measured using a \textit{dispersive qubit readout protocol}. In this protocol, a qubit is coupled to a measurement resonator whose resonance frequency experiences a shift depending on the state of the qubit and by measuring this shift, the state of the qubit is determined~\cite{cqedschoelkopf,cqed,krantz2019quantum}. For superconducting devices, a signal corresponding to the state of a qubit is obtained when a readout pulse is applied to the qubit~\cite{QCSup}. This signal is translated into a single-valued complex number to classify the state of the qubit as ``0" or ``1" using a measurement  \textit{discriminator}. 

\vspace{0.1in}
\noindent{\textbf{Why are measurement operations error prone?}} Measurement errors can be attributed to a variety of factors.

\begin{enumerate}[leftmargin=0cm,itemindent=.5cm,labelwidth=\itemindent,labelsep=0cm,align=left, itemsep=0.1cm, listparindent=0.08cm]

\item The shift in resonance frequency during readout is very sensitive to noise, is device-specific, and drifts in time.  

\item The measurement set-up involves various complex instruments operating across multiple thermal domains that introduces errors due to \textit{crosstalk} and unwanted couplings. The impact of crosstalk is generally not fully understood and considered to be hard to minimize at device level~\cite{khezri2015qubit,QCSup}.

\item Existing discriminators are inefficient and perform poorly for several quantum states~\cite{patel2020disq}.

\item Measurement operations are slow (typically takes about 4-5 microseconds on recent IBMQ hardware and about 800 nanoseconds on Google devices~\cite{ai2021exponential}) and often cause qubits to decay to the ground state during the readout process. 
\end{enumerate}

These factors limit the ability of existing quantum systems to perform fast and accurate qubit measurements at scale. On existing IBMQ and Google hardware, measurement operations tend to be the dominant sources of errors, with median error rates between 2.76\% to 7.1\%, and worst-case error rates as high as 11.7\% to 22.2\%. 

\section{Problem and Motivation}

Measurement operations limit the fidelity of large programs mainly due to two reasons:

\vspace{0.1in}

\begin{enumerate}[leftmargin=0cm,itemindent=.5cm,labelwidth=\itemindent,labelsep=0cm,align=left, itemsep=0.15cm, listparindent=0.3cm]

\item The impact of \textit{measurement crosstalk} increases with the number of measurement operations.

\item \textit{Spatial variation in measurement errors} limit the ability of compilers to avoid the most error-prone physical qubits.

\end{enumerate}

\subsection{Impact of Measurement Crosstalk}
Measurement operations are more vulnerable to errors when a larger number of qubits are simultaneously measured due to measurement crosstalk. For example, the average error rate of simultaneous measurements is 1.26x higher than isolated measurements on Google Sycamore, as shown in Table~\ref{tab:machinespecifications}. 

\begin{table}[htp]
\begin{centering}
\caption{Measurement Errors on Google Sycamore~\cite{arute2019supplementary} }
\setlength{\tabcolsep}{1.8mm} 
\renewcommand{\arraystretch}{1.2}
\label{tab:machinespecifications}
{
\begin{tabular}{ |c|c|c|c|c|} 
\hline
\multirow{2}{*}{Measurement Mode}  & \multicolumn{4}{c|}{Measurement Error Rates (\%)} \\
\cline{2-5}
 & Min & Average & Median & Max \\
\hline 
\hline
Isolated & 2.60 & 6.14 & 5.70 & 11.7 \\
\hline
Simultaneous & 3.30 & 7.73 & 7.10 & 20.9\\
\hline

\end{tabular}
}

\end{centering}
\end{table}

\begin{figure}[htb]
\centering
    \includegraphics[width=\columnwidth]{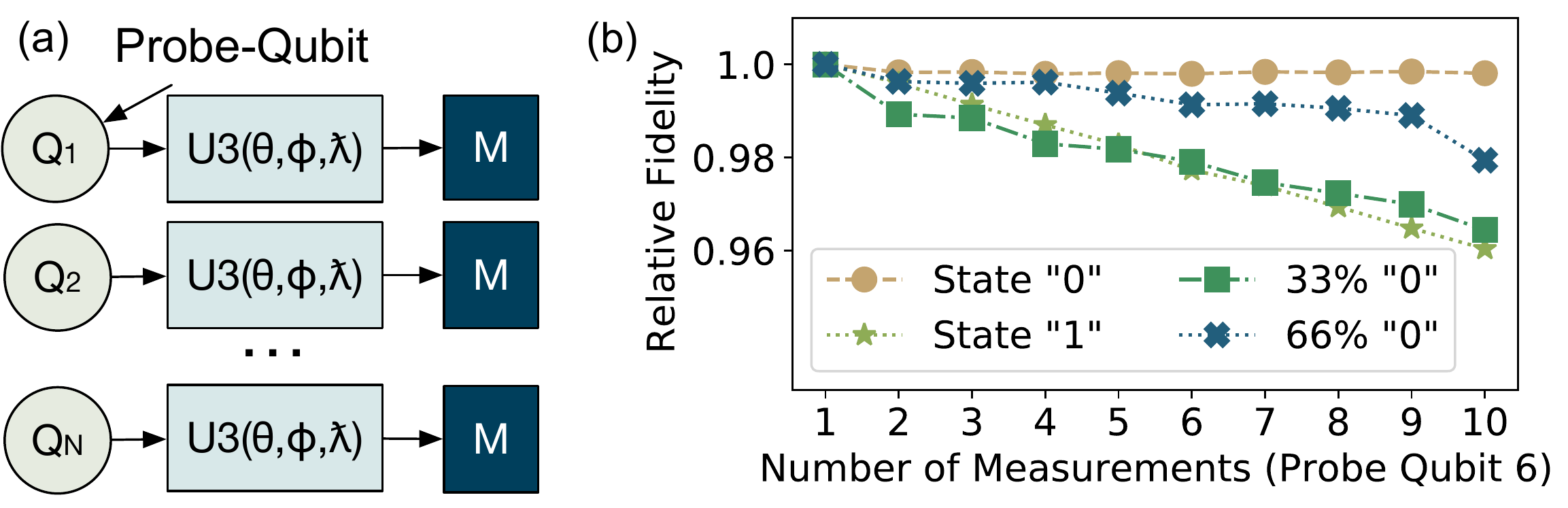}
    \caption{(a) Measurement crosstalk characterization circuits (b) Impact of increasing number of measurements on the Fidelity of the Probe-Qubit.} 
    \label{fig:crosstalkcharacterization}
\end{figure}

\newpage
\noindent \textbf{\textit{Characterization of Measurement Crosstalk:}}\hfill\\
We perform several characterization experiments and observe similar behavior on IBMQ hardware.
To study measurement crosstalk, we perform several characterization experiments on IBMQ hardware. Figure~\ref{fig:crosstalkcharacterization}(a) shows an N-qubit circuit that creates arbitrary quantum states using single-qubit U3 gates~\cite{abhijith2018quantum}. During the experiment, the Probe-Qubit (Q$_1$) is always mapped to the physical qubit on which the impact of measurement crosstalk is being determined, whereas the other N-1 qubits are randomly mapped to the remaining physical qubits of the machine in each sample and generate multiple samples for each N. Note that N denotes the number of measurements and N=1 corresponds to the case when the Probe-Qubit is measured in isolation. For the results shown in Figure~\ref{fig:crosstalkcharacterization}(b), we vary N from 1 to 10 and take 10 samples for each N. 

We compute the mean fidelity of the Probe-Qubit for each N by measuring the \textit{Total Variation Distance (TVD)} between the experimental output and the output from a noise-free quantum computer. Figure~\ref{fig:crosstalkcharacterization}(b) shows the impact of increasing the number of measurements from 1 to 10 for four different quantum states, prepared by specifying the Euler angles for the U3 gates, when Qubit-6 is probed on 27-qubit IBMQ-Paris. We observe that simultaneously measuring a larger number of qubits can reduce the fidelity of these operations significantly. We make similar observations on other qubits and hardware devices. Further experiments also show that the impact of such crosstalk depends on the quantum state and physical qubit and therefore, is hard to characterize. Prior studies state that it is hard to fully understand and minimize measurement crosstalk at device level~\cite{khezri2015qubit,arute2019supplementary}. 

\subsection{Impact of Spatial Variation}
\label{sec:spatialvariation}
To execute a program on a NISQ device, a compiler maps the program qubits onto the physical qubits of the device and translates the high-level program into low-level machine specific instructions. Compilers also need to insert SWAP instructions to overcome the limited connectivity of NISQ devices. Noise-aware compilers~\cite{noiseadaptive,tannu2019not} account for the error characteristics of the underlying quantum hardware and avoid mapping program qubits on to hardware qubits with worst-case errors. However, while this works very efficiently for small programs, spatial variation in measurement error rates often force compilers to map program qubits on unreliable physical qubits, as programs grow in size, because the qubits with the lowest error rates are not spatial neighbors, as shown in Figure~\ref{fig:ibmqtoronto}. For example, it is not possible to map any program with more than six qubits on the 27-qubit IBMQ-Toronto without using a physical qubit with more than 2.7\% measurement error-rate (the median error-rate). Further, the compiler is forced to use physical qubits A and B with more than 20\% measurement error rate for programs with sixteen and twenty-one qubits or more respectively.

\begin{figure}[htb]
\centering
    \includegraphics[width=\columnwidth]{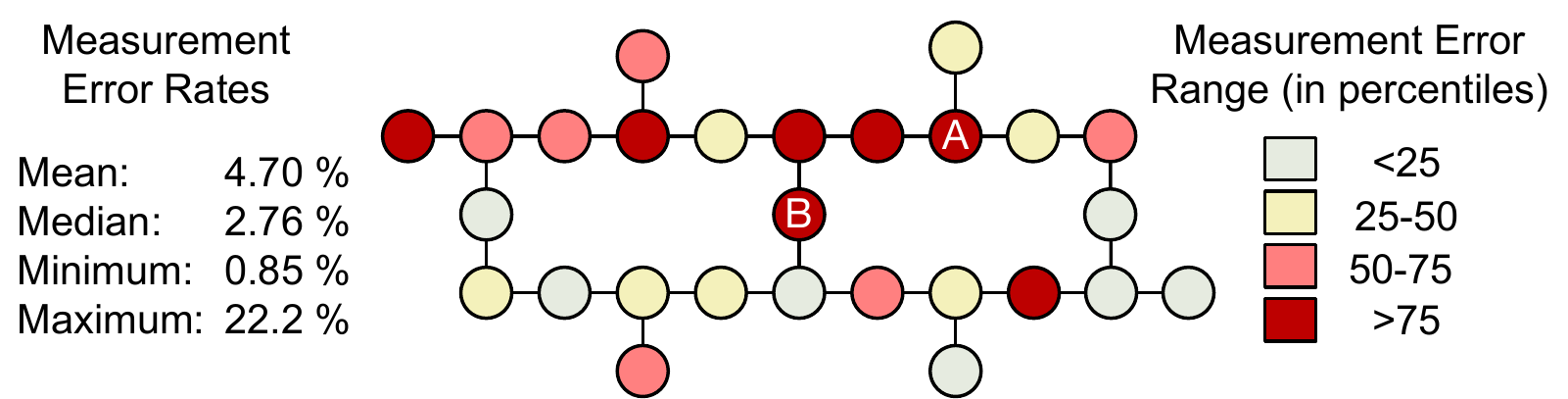}
    \caption{Spatial variation in measurement error rates of qubits on IBMQ-Toronto} 
    \label{fig:ibmqtoronto}
\end{figure}

\subsection{The Insight: Measurement Subsetting}

NISQ applications measure all the program qubits in each trial, subjecting all the trials to the accumulated measurement errors from all the qubits. We can reduce the impact of measurement errors on the fidelity of NISQ programs by:
{\it{
\begin{enumerate}[leftmargin=0cm,itemindent=.5cm,labelwidth=\itemindent,labelsep=0cm,align=left, itemsep=0.15cm, listparindent=0.4cm]
    \item Reducing the number of measurement operations by performing some trials with measurements only on a subset of qubits and effectively lowering the impact of crosstalk. 
    
    \item Remapping to ensure that the subset measurements are performed on the qubits with the lowest error rates. This allows the us to get an effective measurement error rate that is closer to the minimum rather than the average.
\end{enumerate}
}}


\vspace{0.05 in}
\noindent \textbf{Correlation versus Fidelity Trade-off in Subsetting}: \hfill \\
Quantum computers obtain their exponential power by creating  and manipulating highly correlated states. To measure this correlated state, a NISQ program measures all the qubits in each trial. If there was no correlation, and each qubit had an independent probability of being in the $0$ or $1$ state, then one could simply split the trials into N groups (one group for each qubit), obtain the independent probability of being in $0$ or $1$ state for each qubit, and then obtain the probability distribution over all the qubits through multiplication. While this approach has high fidelity for each trial (only one measurement), it captures zero correlation between the qubits. Measuring a subset of qubits (larger than a single qubit but not all qubits) captures some correlation (within the measured qubits) but the correlations between these marginal distributions remain unknown, and therefore, multiplication or a tensor product may not yield the correct output distribution.

\subsection{The Goal: Reducing Measurement Errors}
Thus, measuring all the qubits provides full correlation (but low fidelity) and fewer measurements provide higher fidelity (but weaker correlation). Ideally, we want both full correlation and high fidelity. The goal of this paper is to design scalable and effective policies that can improve application fidelity by retaining the global correlation of the original program, while simultaneously benefiting from the higher fidelity obtained from fewer measurements. Next, we discuss our proposed design, JigSaw.
\begin{figure*}[htb]
\centering
  \includegraphics[width=\linewidth]{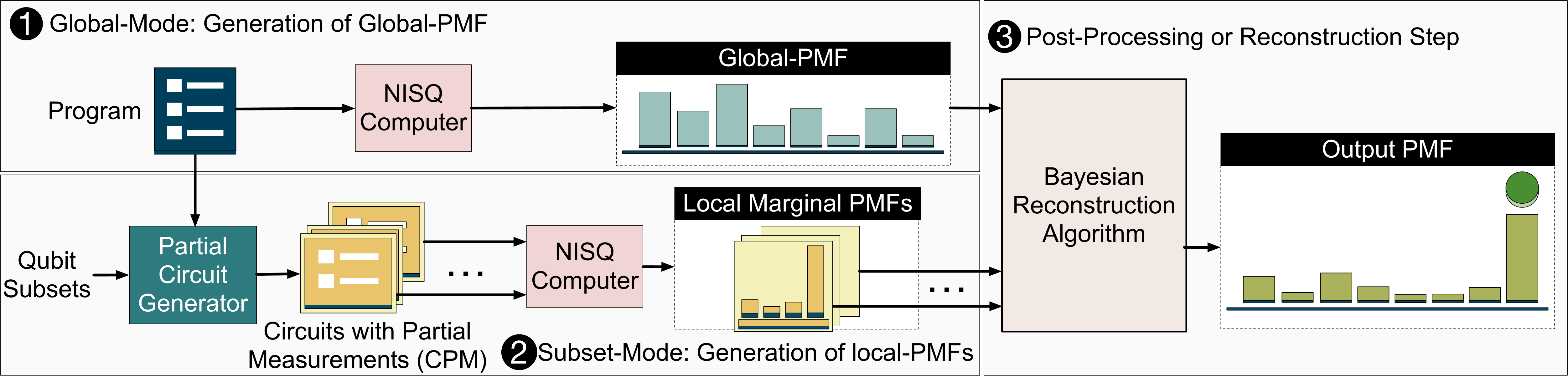}
  \caption{Overview of JigSaw. JigSaw runs the program in two modes: global-mode (all program qubits are measured) and subset-mode (only a subset of qubits are measured), and then uses the local-PMFs generated in subset-mode to update the global-PMF obtained during the execution in global-mode.}
\label{fig:partialdesignoverview}
\end{figure*}
\newpage
\section{JigSaw: Overview and Design}
\label{sec:measurementsubsetting}

We propose {\em JigSaw}, a framework that relieves the NISQ program from requiring to measure all the qubits of a program in each trial. Figure~\ref{fig:partialdesignoverview} shows the overview of JigSaw. JigSaw executes a program in two modes. First, JigSaw executes half of the trials in the \textit{global-mode} in which a program is executed in its entirety and all the qubits are measured to obtain the global probability mass function (PMF).\footnote{We use Probability Mass Function (PMF) for the results of program execution because these include discrete and not continuous values.} Second, for the remaining trials, JigSaw runs additional copies of the program or \textit{Circuits with Partial Measurements (CPM)} that measure fewer qubits in the \textit{subset-mode}, to obtain more accurate marginal or local-PMFs over the  \textit{subset} of qubits being measured. However, CPM alone cannot be used to infer the output PMF of a program  without information about the correlations between these local-PMFs. To address this challenge, JigSaw employs a post-processing or \textit{reconstruction} step that updates the global-PMF using the local-PMFs. This enables JigSaw to improve the application fidelity while simultaneously retaining the global correlation without requiring any additional trials. 

\subsection{Global-Mode: Generation of Global-PMF}
In this mode, JigSaw executes the entire program and measures all the program qubits to produce a global-PMF. This is identical to the baseline policy and is done for half of the trials. We use Noise-Aware SABRE~\cite{li2018tackling} for compilation to obtain a global-PMF with high fidelity. A NISQ compiler maps the logical qubits of a program on to the physical qubits of the hardware and generates a schedule by translating the high-level instructions into low-level machine specific operations. To overcome limited connectivity on NISQ hardware, compilers also insert \texttt{SWAP} instructions to bring two non-adjacent qubits next to each other, so that \texttt{CNOT} operations can be performed between them. Noise-Aware SABRE accounts for the hardware error characteristics and generates a schedule that maximizes the \textit{Expected Probability of Success} (EPS)~\cite{nishio}. EPS is the expected probability of successfully executing each gate and measurement operation in a schedule and is computed at compilation time by using the error rates obtained from the daily calibration report of the NISQ hardware.

\subsection{Subset-Mode: Generation of Local-PMFs}
In the subset-mode, JigSaw runs several Circuits with Partial Measurements (CPM), for the remaining half of the trials which are equally distributed between the CPM. We discuss how CPM are generated and optimized for greater fidelity. 

\subsubsection{Circuits with Partial Measurements} 
\begin{figure}[htb]
\centering
    \includegraphics[width=1.0\columnwidth]{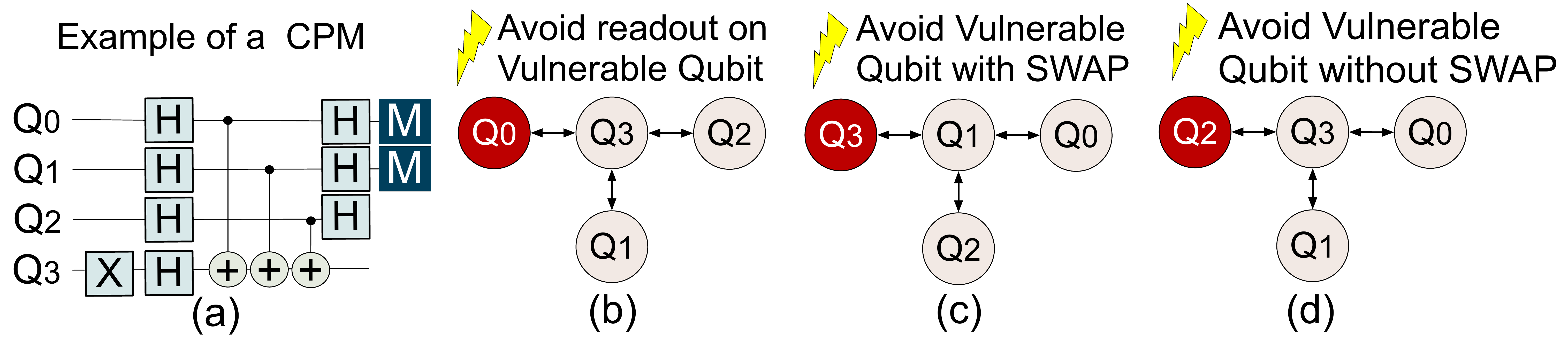}
    \caption{(a) Example of a CPM (b) Compiler avoids vulnerable qubit. Mapping that avoids vulnerable qubit (c) with extra \texttt{SWAP} (d) without incurring extra \texttt{SWAP}} 
    \label{fig:compileroptimizations}
\end{figure}

\hfill \\ 
\noindent A CPM is identical to the original program, except that it measures only a subset of qubits. CPM produces high-fidelity local-PMFs over the qubits measured. For example, Figure~\ref{fig:compileroptimizations}(a) shows a CPM of a BV-4 program that measures 2 qubits, $\textrm{Q}_0$ and $\textrm{Q}_1$. The key parameter in JigSaw is the number of qubits measured in a CPM and is called the \textit{subset size}. Our default design uses CPM that measure 2 qubits. This is the smallest possible subset size that captures some correlation while performing fewest possible measurements. Measuring only one qubit in a CPM captures zero correlation and therefore, not used. By default, we use a sliding window method to generate the CPM so that we get N unique CPM for an N-qubit program. For example, for a $4$-qubit program with qubits $\textrm{q}_0,\textrm{q}_1,\textrm{q}_2,\textrm{q}_3$, we generate $4$ CPM,   measuring $(\textrm{q}_0,\textrm{q}_1), (\textrm{q}_1,\textrm{q}_2), (\textrm{q}_2,\textrm{q}_3)$, and $(\textrm{q}_0,\textrm{q}_3)$. Therefore, the number of CPM is same as the number of qubits. 

\begin{figure*}[htb]
\centering
    \includegraphics[width=0.9\linewidth]{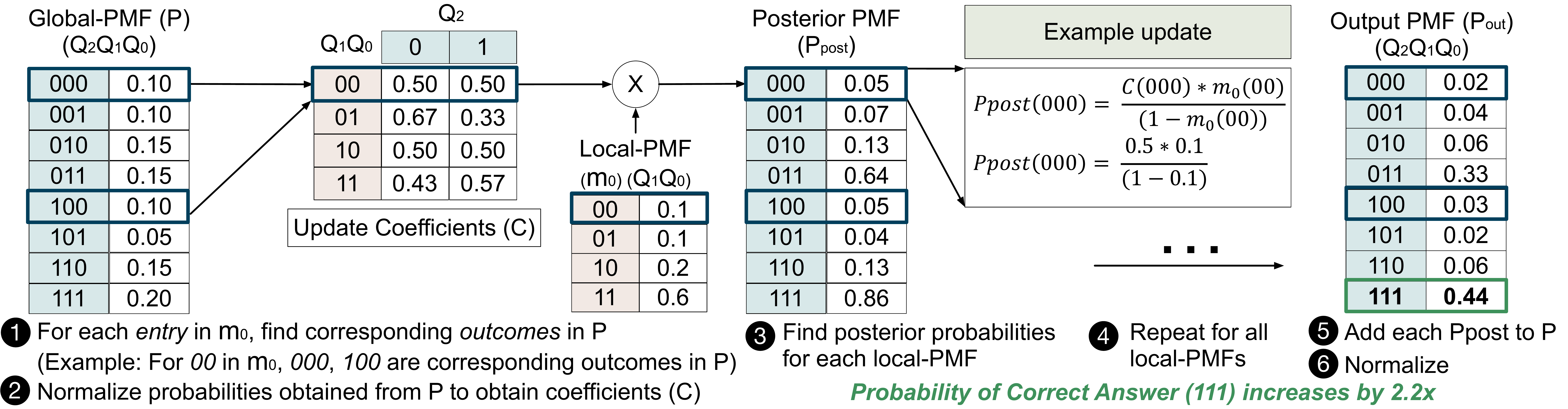}
    \caption{Steps in Bayesian Reconstruction for a 3-qubit program (qubits $\textrm{Q}_2$, $\textrm{Q}_1$, $\textrm{Q}_0$) using a CPM measuring $\textrm{Q}_1$, $\textrm{Q}_0$.}
    \label{fig:examplebayesianupdate}
\end{figure*}

\subsubsection{Optimizations to Improve Fidelity of the CPM} \hfill \\
As JigSaw heavily relies on the accuracy of the local-PMFs, the compiler recompiles each CPM to maximize their fidelity.

\vspace{0.05in}
\noindent \textbf{Optimizing for Measurement Errors}: We recompile each CPM to exploit variability in measurement errors and ensure that measurements of desired program qubits are performed on the strongest physical qubits. The compiler avoids measurements on the physical qubit(s) with the highest readout error rate, referred to as \textit{vulnerable qubit(s)}, in an allocation. For example, the compiler eliminates the allocation in Figure~\ref{fig:compileroptimizations}(b) for the CPM in Figure~\ref{fig:compileroptimizations}(a) to avoid   readout of $\textrm{Q}_0$ on the vulnerable qubit and selects an alternate allocation. 

\vspace{0.05in}
\noindent \textbf{Avoiding Extra SWAPs}: To avoid measurement on the vulnerable qubit(s), the compiler often uses alternate qubit allocations, some of which may need insertion of extra \texttt{SWAP} instructions. However, we avoid such allocations that require extra \texttt{SWAP}s to avoid additional gate errors. For example, the compiler selects the allocation shown in Figure~\ref{fig:compileroptimizations}(d) over Figure~\ref{fig:compileroptimizations}(c) because the latter requires an extra \texttt{SWAP} (\texttt{SWAP} $\textrm{Q}_0$, $\textrm{Q}_1$). While in most cases, the compiler finds alternate mappings without incurring extra \texttt{SWAP}s, for cases where the compiler cannot not find a mapping without inserting an extra \texttt{SWAP}, it picks the mapping that maximizes the EPS. 

The ability to find alternative mappings depends on device connectivity, spatial location of good qubits, and program characteristics. Our studies show most CPM can reuse qubit allocations chosen by the Ensemble of Diverse Mappings~\cite{micro1} policy. Also, we use SABRE, which has low latency.

\subsection{Post-processing via Bayesian Method}
\label{sec:basicreconstruction}
JigSaw produces (N+1) PMFs for a program with N qubits: one global-PMF and one local-PMF for each of the N CPMs. The post-processing step aims to combine the higher fidelity local-PMFs from each CPM into the global-PMF. To this end, we propose {\em Bayesian Reconstruction} algorithm, which is inspired by Bayesian updating in statistics, whereby a prior probability estimate is updated using additional information~\cite{joyce2003bayes}. For JigSaw, the global-PMF ($\textrm{P}$) offers the prior probability estimate, whereas the set of marginals (\textrm{M}) obtained from the CPM provides the additional information. We use the term \textit{marginal} to denote a set comprising of a subset of qubits measured in a CPM and the local-PMF produced by the CPM. 

Consider an example of a 3-qubit program. Let $\textrm{P} = \{000:\textrm{a}, 010:\textrm{b}, 010:\textrm{c}, 011:\textrm{d}, 100:\textrm{e}, 101:\textrm{f}, 110:\textrm{g}, 111: \textrm{h}\}$ represent a generic global-PMF of a $3$-qubit program (with qubits $\textrm{Q}_2, \textrm{Q}_1, \textrm{Q}_0$), where $\textrm{a}$ to $\textrm{h}$ are the probabilities of observing outcomes $000$ to $111$ respectively. Similarly, $\textrm{m}_\textrm{0} = [\{00:\alpha, 01:\beta, 10:\gamma, 11:\delta\},[1,0]]$ refers to a generic marginal for a CPM measuring qubits $[\textrm{Q}_\textrm{1},\textrm{Q}_\textrm{0}]$.

The steps for the Bayesian Reconstruction algorithm are described in Algorithm~\ref{alg:bayesianupdate} (Appendix). The algorithm uses each marginal to update the probabilities of each outcome in the global-PMF ($\textrm{P}$). The algorithm starts by searching for all outcomes in $\textrm{P}$ associated with each outcome in a marginal, by evaluating the bits at the corresponding qubit positions. For example, for $(\textrm{Q}_\textrm{1},\textrm{Q}_\textrm{0}) = 00$ in the marginal $\textrm{m}_\textrm{0}$, the candidates in $\textrm{P}$ are $(\textrm{Q}_\textrm{2},\textrm{Q}_\textrm{1},\textrm{Q}_\textrm{0}) = 000$ and $100$. For each marginal in \textrm{M}, the Bayesian Update function generates an updated PMF ($\mathrm{P_{post}}$) by updating the probabilities of each outcome in $\textrm{P}$ using the associated probabilities in the marginals. For example, the probabilities of $000$ and $100$, i.e., $\textrm{a}$ and $\textrm{d}$ respectively, are updated in proportion with $\alpha$ (corresponding to $00$ in m$_0$). The algorithm produces a posterior output PMF ($\mathrm{P_{out}}$) by adding all the intermediate PMFs ($\mathrm{P_{post}}$) to the global-PMF ($\mathrm{P}$). The algorithm is recursively called and terminates when the Hellinger Distance~\cite{hellingerdistance} between the output PMF, $\mathrm{P_{out}}$, prior to and post the function call does not change, implying the two PMFs are similar.

We explain the steps involved in the Bayesian update using a quantitative example.  
Figure~\ref{fig:examplebayesianupdate} shows the update sequence using experimental data for a 3-qubit program whose global-PMF is denoted by $\mathrm{P}$ $($in order $\textrm{Q}_2, \textrm{Q}_1, \textrm{Q}_0)$ using a marginal $\textrm{m}_0$ from a CPM that measures $(\textrm{Q}_1,\textrm{Q}_0)$.

\vspace{-0.05 in}

\begin{itemize}[leftmargin=0cm,itemindent=.5cm,labelwidth=\itemindent,labelsep=0cm,align=left, itemsep=0.02cm, listparindent=0.4cm]
\setlength{\parskip}{0pt}
    \item \textbf{Step 1} : The algorithm searches for all candidate outcomes in $\mathrm{P}$ for each entry in marginal $\textrm{m}_0$ by evaluating the bits at the corresponding qubit positions. For example, $000$ and $100$ are the candidate outcomes in $\mathrm{P}$ for $00$ in $\textrm{m}_0$, obtained by matching the values for $(\textrm{Q}_1, \textrm{Q}_0)$ in $\mathrm{P}$ and $\textrm{m}_0$.   
    \item \textbf{Step 2} : Next, the function computes the Update Coefficients $\mathrm{C}$ for each of the outcomes in $\mathrm{P}$ by normalizing their respective probabilities of occurrence in $\mathrm{P}$.  
    \item \textbf{Step 3} : Next, the algorithm computes the posterior probabilities for each observed outcome in $\mathrm{P}$ by scaling them using the corresponding probabilities observed in $\textrm{m}_0$. Figure~\ref{fig:examplebayesianupdate} explicitly shows the computation for obtaining the posterior probability of outcome $000$ ($\mathrm{P_{post}}[000]$) using the Update Coefficient $\mathrm{C}[000]$ and marginal information $\textrm{m}_0[00]$. 
    \item \textbf{Step 4} : The algorithm repeats Steps 1-3 to generate an intermediate posterior PMF ($\mathrm{P_{post}}$) for each marginal.
    \item \textbf{Step 5} : Each $\mathrm{P_{post}}$ is added to the global-PMF ($\mathrm{P}$).
    \item \textbf{Step 6} : The final output PMF ($\mathrm{P_{out}}$) is obtained by normalizing the probabilities. 
    
\end{itemize}

For readability, Figure~\ref{fig:examplebayesianupdate} only shows the steps for marginal [$\textrm{Q}_1, \textrm{Q}_0$], whereas the output PMF shown is obtained from recursive updates with additional marginals. Note that as the Bayesian updates for each CPM are performed independently and the intermediate PMFs are added in the final step, the order of updates does not matter. 

\subsection{Multi-Layer JigSaw (JigSaw-M)}
Our studies show that the performance of JigSaw saturates when additional CPM of the same subset size that do not offer any incremental information are used. However, we can design more unique CPM by using different subset sizes and improve the application fidelity even further.

\subsubsection{Global and Subset-Modes for JigSaw-M}  \hfill \\ 
The global mode for JigSaw-M is identical to JigSaw that generates the global-PMF by executing the program and measuring all the qubits. 
The subset-mode for JigSaw-M executes CPM of non-uniform subset sizes $\textrm{s}$, such that $\textrm{s}_{\textrm{min}} \leq \textrm{s} \leq \textrm{s}_{\textrm{max}}$, where $\textrm{s}_{\textrm{max}}$ and $\textrm{s}_{\textrm{min}}$ are the maximum and minimum subset sizes respectively. 
We use a sliding window method to generate unique CPM for each subset size, similar to JigSaw, but other methods can be used too. If CPM of $\mathbb{S}$ different sizes are used, JigSaw-M produces ($\mathbb{S}$N+1) PMFs for an N-qubit program, one global-PMF and N local-PMFs for each subset size. By default, our design uses CPM of sizes 2 to 5.

\subsubsection{Adapting Reconstruction for JigSaw-M}  \hfill \\ 
JigSaw-M comprises of CPM of $\mathbb{S}$ different sizes. There is a choice between which CPM must be used first to update the global-PMF. However, note that there exists a \textit{trade-off} between the fidelity of a CPM and the correlation it can capture depending on the number of qubits measured in the CPM. 
A smaller CPM offers higher fidelity due to fewer measurement errors but captures limited correlation. Alternately, a larger CPM offers higher correlation, but has lower fidelity since it is more prone to measurement errors. Thus, for JigSaw-M, the reconstruction algorithm first updates the global-PMF using the CPM of the highest size ($\textrm{s}_{\textrm{max}}$), limiting the loss of global
correlation. The updated PMF ($\textrm{P}_{\textrm{s}}$) is then further enhanced using information from CPM of the next higher size. The process is repeated until the smallest CPM are used.  This top-down ordering maximally preserves the global correlation, while simultaneously improving the fidelity.

We discuss the evaluation methodology before discussing the impact of JigSaw on the fidelity of NISQ applications.

\section{Evaluation Methodology}
\label{sec:evaluation}

\subsection{Quantum Hardware Platforms}
For all our evaluations, we use three different quantum computers from IBM: $27$-qubit {\em IBMQ-Toronto}, $27$-qubit {\em IBMQ-Paris}, and $65$-qubit {\em IBMQ-Manhattan}.  

\subsection{Baseline Compiler}
For the baseline, we use Noise-Aware SABRE~\cite{li2018tackling} to compile and map the program onto the physical qubits with the lowest error rates. We also evaluate JigSaw against an Ensemble of Diverse Mappings (EDM) policy that runs independent copies of the program on different groups of physical qubits~\cite{micro1} and improves the ability to infer the
correct answer. Note that while we use Noise-Aware SABRE, other noise-adaptive compilers~\cite{noiseadaptive,tannu2019not} may be used too. 

\subsection{Benchmarks} 
We use the benchmarks described in Table~\ref{tab:benchmarks}. The type and size of benchmarks are derived from prior works~\cite{li2018tackling,tannu2019not,noiseadaptive,micro1}. Note that the IBMQ machines used for evaluations have a Quantum Volume~\cite{quantumvolume} of 32, which means that square circuits of only up to size 5 can be run reliably and therefore, the size of the benchmarks used is already much larger even if they do not use all the qubits that are present on the machine. 

\begin{table}[htb]
\begin{center}
\begin{small}
\caption{Details of NISQ benchmarks}
\setlength{\tabcolsep}{1.0mm} 
\renewcommand{\arraystretch}{1.2}
\label{tab:benchmarks}
{
\begin{tabular}{ |c|c|c|c|c| } 
\hline
\multirow{2}{*}{Name} & \multirow{2}{*}{Algorithm} & \#Qubits  & 1Q & 2Q \\
 & &  (n) & Gates & Gates \\
\hline 
\hline
BV-n & Bernstein-Vazirani~\cite{bernsteinvazirani} & 6 & 2(n+1) & n \\
\hline
Graycode-n & Graycode Decoder & 18 & n/2& (n-1)\\
\hline
QAOA-n (p=1) & Maxcut with p=1~\cite{qaoa} & 8 & 4n & (n-1) \\
\hline
QAOA-n (p=2) &  Maxcut with p=2 & 10, 14 & 6n & 2(n-1) \\
\hline
QAOA-n (p=4) &  Maxcut with p=4 & 10, 12 & 10n & 4(n-1)\\
\hline
\multirow{2}{*}{GHZ-n} & Greenberger-Horne& \multirow{2}{*}{14} & \multirow{2}{*}{1} & \multirow{2}{*}{(n-1)}\\
& -Zeilinger~\cite{GHZ} & & & \\
\hline
Ising-n & Ising model~\cite{ising1925beitrag} & 10 & n(4.5n-2) & n(n-1)\\
\hline

\end{tabular}}
\end{small}
\end{center}
\end{table}

\subsection{Experimental Setup: Number of Trials}

\begin{figure}[htp]
\centering
    \includegraphics[width=1.0\columnwidth]{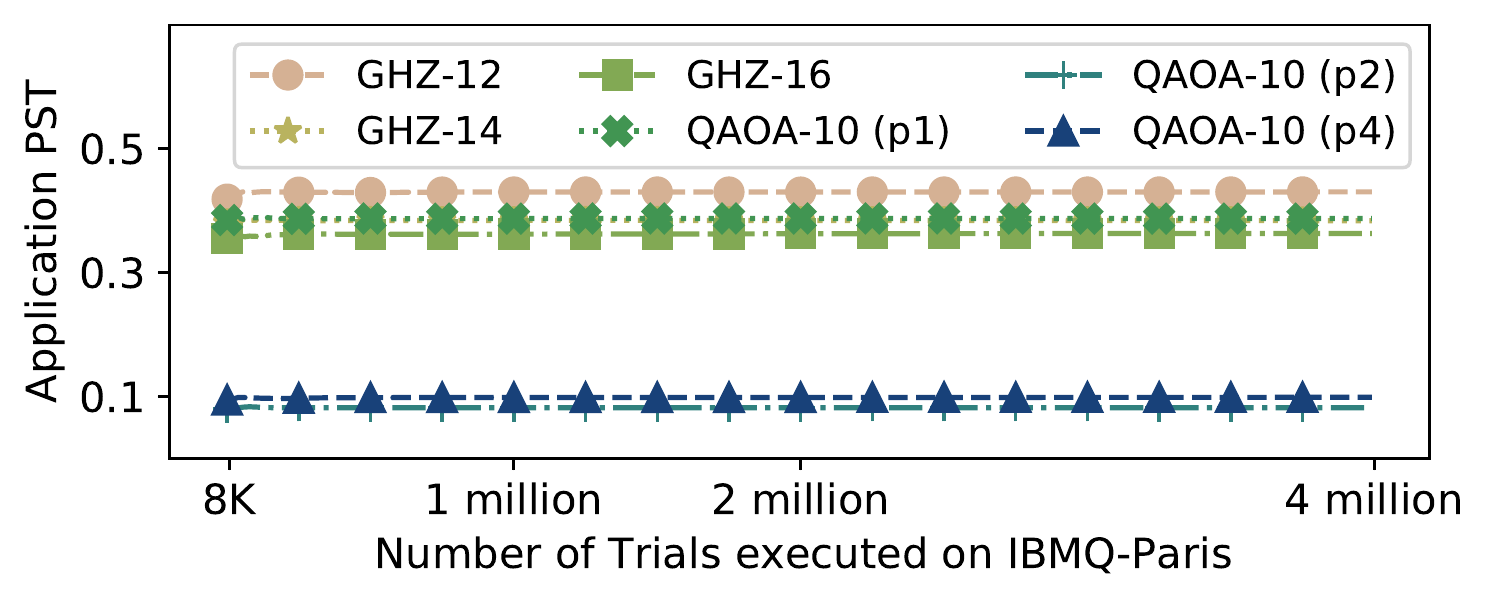}
    \caption{Impact of Number of Trials on Probability of Successful Trial (PST) of Applications.}
    \label{fig:pstsaturation}
\end{figure}

We use between 32K to 256K trials for the baseline depending on program size. This represents the highest fidelity that can be obtained by increasing trials alone and serves as a strong baseline as more trials do not improve fidelity (mainly due to correlated errors).  Figure~\ref{fig:pstsaturation} shows the Probability of Successful Trial (PST) of several GHZ and QAOA benchmarks executed on IBMQ-Paris for up to 4 million trials. We observe similar behavior for other workloads and machines. 

\begin{figure*}[htb]
\centering
    \includegraphics[width=\textwidth]{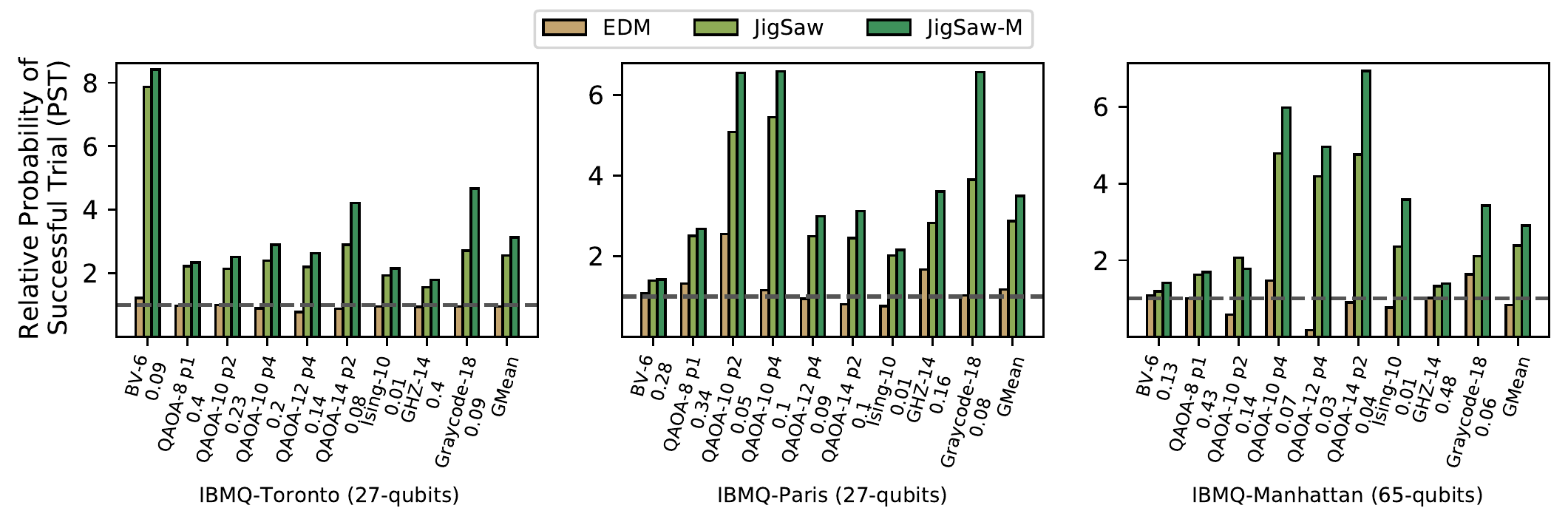}
    \caption{Probability of Successful Trial (PST) from JigSaw and JigSaw-M relative to baseline and comparison with prior work Ensemble of Diverse Mappings (EDM)~\cite{micro1}. Number below the label shows absolute PST for the benchmark.} 
    \label{fig:hierarchypst}
\end{figure*}

For EDM~\cite{micro1}, we use an ensemble of four mappings and the trials are equally divided among the mappings. For JigSaw and JigSaw-M, the trials are equally split between the global-mode and the subset-mode. In the subset-mode the trials are equally split between all the CPM for both JigSaw and JigSaw-M.  Therefore, JigSaw, JigSaw-M, and EDM all use the {\bf same number of trials} as the baseline. We run all the experiments within the same calibration cycle but observe similar results across different calibration cycles. We use equal split for simplicity because the fidelity saturates for the number of trials used in our evaluation. If the number of trials is severely limited, the split between global-mode and subset-mode can be tuned to possibly obtain even larger gains. We provide an estimate of the number of trials required for each CPM in Appendix~\ref{sec:trials}.

\subsection{Figure-of-Merit}
At present, there is no standard single metric to evaluate NISQ application fidelity. Therefore, for our evaluations, we study three generic metrics and an application-specific figure-of-merit for the QAOA benchmarks. These metrics are derived from prior works and the details are discussed below:

\vspace{0.1in}
\noindent \textbf{(1)} \textit{\textbf{Probability of Successful Trial (PST)}}~\cite{tannu2018case,tannu2019not,micro1,liureliability,micro3, noiseadaptive} is the ratio of the number of trials with the correct output to the total number of trials, as described in Equation~\eqref{eq:pst}.
\begin{equation}
    \label{eq:pst}
    PST = \frac{\textrm{Number of trials with the correct output}}{\textrm{Total number of trials}}
\end{equation}

\vspace{0.1in}
\noindent \textit{\textbf{(2) Inference Strength (IST)}}~\cite{micro1,liureliability,patel2020veritas} is used to quantify the ability to infer the solution and distinguish it from incorrect outcomes. IST is defined as the ratio of the probability of occurrence of the correct outcome to the probability of occurrence of the most frequent erroneous outcome, as shown in Equation~\eqref{eq:ist}.
\begin{equation}
    \label{eq:ist}
    IST = \frac{\textrm{Probability of correct outcome}}{\textrm{Probability of most frequent incorrect outcome}}
\end{equation}

\vspace{0.1in}
\noindent \textit{\textbf{(3) Fidelity}} of a program is obtained by measuring the Total Variation Distance (TVD)~\cite{TVD} between the output distributions on a noise-free quantum computer (P) and real hardware (Q). TVD allows us to measure the fidelity of quantum programs~\cite{sanders2015bounding} whose output can be a probability distribution with more than one correct answer. The Fidelity ranges between 0 and 1, where 1 represents two identical distributions and 0 means completely
dissimilar distributions, as shown in Equation~\eqref{eq:corr}. While we use TVD, Hellinger Distance~\cite{hellingerdistance,micro3} or Kullback–Leibler divergence~\cite{kullback1997information} may be used too and these metrics are closely related~\cite{TVD}.

\begin{equation}
\begin{split}
TVD(P,Q) = \sum_{i=1}^{k}\mid \mid P_i - Q_i \mid \mid \\
Fidelity(P,Q) = 1- TVD(P,Q)
\end{split}\label{eq:corr}
\end{equation}

\vspace{0.1in}
\noindent
\textit{A higher Probability of Successful Trial (PST), Fidelity, and Inference Strength (IST) is desirable. }

\vspace{0.1in}
\noindent \textbf{\textit{(4) Approximation Ratio Gap (ARG)}}~\cite{alam2020circuit} is 
an application-specific metric for \textit{Quantum Approximate Optimization Algorithm (QAOA)~\cite{qaoa}} benchmarks. To solve MaxCut problems with QAOA, the classical cost function which must be maximized is translated into a cost Hamiltonian and the goal is to maximize the expectation value of the cost Hamiltonian. The expectation value of a cost function is computed by taking a mean over the samples in the output distribution of the QAOA circuit. The Approximation Ratio (AR)
is defined as the ratio between the mean cost function value
over these samples and the actual maximum function value of the optimal solution and is used to quantify QAOA performance~\cite{crooks2018performance,zhou2020quantum}. The Approximation Ratio Gap (ARG) denotes the percentage difference between the  approximation ratio obtained on an ideal quantum computer (AR$_\textrm{ideal}$) and real hardware (AR$_\textrm{real}$), as described in Equation~\eqref{eq:arg}. \textit{A lower ARG is desired as it indicates a
performance closer to the noise-free scenario.}
\begin{equation}
\label{eq:arg}
{ARG = \frac{100*(\textrm{AR}_\textrm{ideal}- \textrm{AR}_\textrm{real})}{\textrm{AR}_\textrm{ideal}}}
\end{equation}

\section{Results and Sensitivity Studies}
\label{sec:finalresults}
In this Section, we discuss the impact of JigSaw on the reliability of NISQ applications.

\subsection{Results for Probability of Successful Trial}
Figure~\ref{fig:hierarchypst} shows the improvement in Probability of Successful Trial (PST) using JigSaw and JigSaw-M. Our evaluations using three different quantum hardware from IBM and tens of quantum benchmarks show that JigSaw the PST by 2.91x on average and by up-to 7.87x compared to the baseline. JigSaw-M improves PST by 3.65x on average and up-to 8.42x compared to the baseline. Compared to JigSaw, JigSaw-M improves the PST of applications by 1.26x on average and up-to 1.72x. 

\subsection{Results for Inference Strength}
Inference Strength (IST) determines the capability to suppress correlated errors and infer the correct answer of a program from the the output PMF.  Table~\ref{tab:inferencestrength} 
shows the improvement in IST for EDM, JigSaw, and JigSaw-M relative to the baseline. Note that the average here is the geometric mean. JigSaw improves the IST on average by 2.19x and up-to 21.7x compared to the baseline, whereas JigSaw-M improves the IST on average by 2.82x and up-to 27.9x. Unlike EDM which only improves the IST, JigSaw improves both PST and IST. 

\begin{table}[htb]
\begin{center}
\begin{small}
\caption{
Inference Strength (IST) obtained from EDM, JigSaw, and JigSaw-M relative to the Baseline}

\setlength{\tabcolsep}{1.0mm} 
\renewcommand{\arraystretch}{1.1}
\label{tab:inferencestrength}
{
\begin{tabular}{ |c||c|c|c||c|c|c||c|c|c|} 
\hline
IBMQ & \multicolumn{3}{c||}{EDM} & \multicolumn{3}{c||}{JigSaw} & \multicolumn{3}{c|}{JigSaw-M}  \\
\cline{2-10}
(Hardware) & Min & Max & Avg & Min & Max  & Avg & Min & Max & Avg \\
\hline
\hline
Toronto & 0.92 & 2.25 & 1.36 & 1.22 & 21.7 & 2.87 & 1.23 & 27.9 & 3.84 \\
 \hline 

Paris & 0.78 & 6.54 & 1.36 & 1.07 & 9.07 & 2.33 & 1.09 & 28.1 & 3.13 \\
 \hline 

Manhattan & 0.75 & 2.74 & 1.27 & 0.81 & 3.12 & 1.35 & 0.83 & 3.40 & 1.46 \\
 \hline 

\end{tabular}}

\end{small}
\end{center}
\end{table}

\subsection{Results for Fidelity}
Table~\ref{tab:hdist} compares the Fidelity for EDM, JigSaw, and JigSaw-M relative to the baseline. For instance, on IBMQ-Toronto, EDM degrades Fidelity by 0.96x on an average, whereas JigSaw and JigSaw-M improve it by 2.17x and 2.54x respectively. Overall, JigSaw improves the Fidelity on average by 2.12x, whereas JigSaw-M improves the Fidelity on average by 2.47x and by up-to 8.41x compared to the baseline.  Therefore, the output distributions of programs obtained from JigSaw and JigSaw-M have higher Fidelity and are significantly more similar to the distributions obtained on a noise-free quantum computer.

\begin{table}[htb]
\begin{center}
\begin{small}
\caption{
{Fidelity obtained from EDM, JigSaw, and JigSaw-M relative to the baseline}}

\setlength{\tabcolsep}{0.9mm} 
\renewcommand{\arraystretch}{1.25}
\label{tab:hdist}
{
\begin{tabular}{ |c||c|c|c||c|c|c||c|c|c|} 
\hline
IBMQ & \multicolumn{3}{c||}{EDM} & \multicolumn{3}{c||}{JigSaw} & \multicolumn{3}{c|}{JigSaw-M}  \\
\cline{2-10}
(Hardware) & Min & Max & Avg & Min & Max  & Avg & Min & Max & Avg \\
\hline
\hline
Toronto & 0.78 & 1.22 & 0.96 & 1.07 & 7.86 & 2.17 & 1.07 & 8.41 & 2.54 \\
 \hline 

Paris & 0.77 & 2.54 & 1.19 & 1.09 & 5.07 & 2.33 & 1.11 & 6.52 & 2.77 \\
 \hline 

Manhattan & 0.43 & 1.62 & 0.93 & 1.18 & 3.26 & 1.84 & 1.28 & 4.43 & 2.10 \\
 \hline

\end{tabular}}

\end{small}
\end{center}
\end{table}
\subsection{Results for Approximation Ratio Gap}
We use Approximation Ratio Gap (ARG) as an application-specific metric for the QAOA benchmarks. 
A lower ARG is desired as it indicates a performance closer to the noise-free scenario~\cite{alam2020circuit}. Table~\ref{tab:eev_comparison} compares the ARG for the QAOA benchmarks evaluated in this paper. We observe that JigSaw reduces the ARG by 0.41x on average and by up-to 0.14x compared to the baseline. Alternately, JigSaw-M reduces the ARG by 0.31x on average and up-to 0.08x compared to the baseline. Overall, JigSaw and JigSaw-M consistently outperforms the baseline and EDM. Overall, JigSaw outperforms the baseline and prior work, EDM~\cite{micro1}, across all the four metrics. 

\begin{table}[htb]
\begin{center}
\begin{small}
\caption{
{Comparison of Approximation Ratio Gap (ARG) (values are \%) between Baseline, EDM, JigSaw, and JigSaw-M}}

\setlength{\tabcolsep}{1.2mm} 
\renewcommand{\arraystretch}{1.2}
\label{tab:eev_comparison}
{
\begin{tabular}{ |c||c||c|c|c|c|} 
\hline
Machine & Workload & Baseline & EDM & JigSaw & JigSaw-M \\
\hline
\hline
\multirow{5}{*}{Tor} & QAOA-8 p1 & 19.6 & 19.4 & 2.83 & 1.59 \\ \cline{2-6} 
 & QAOA-10 p2 & 24.5 & 24.0 & 12.3 & 10.6 \\ \cline{2-6} 
 & QAOA-10 p4 & 23.4 & 24.3 & 10.5 & 8.50 \\ \cline{2-6} 
 & QAOA-12 p4 & 12.3 & 13.8 & 4.82 & 3.11 \\ \cline{2-6} 
 & QAOA-14 p2 & 9.86 & 9.74 & 4.06 & 2.48 \\\hline 

\multirow{5}{*}{Par} & QAOA-8 p1 & 21.7 & 17.5 & 3.91 & 2.50 \\ \cline{2-6} 
 & QAOA-10 p2 & 35.0 & 29.2 & 19.0 & 16.3 \\ \cline{2-6} 
 & QAOA-10 p4 & 30.3 & 29.4 & 8.60 & 6.19 \\ \cline{2-6} 
 & QAOA-12 p4 & 10.5 & 11.8 & 5.63 & 4.98 \\ \cline{2-6} 
 & QAOA-14 p2 & 8.50 & 9.66 & 3.95 & 2.95 \\\hline 

\multirow{5}{*}{Man} & QAOA-8 p1 & 18.2 & 18.4 & 8.87 & 8.23 \\ \cline{2-6} 
 & QAOA-10 p2 & 28.1 & 31.9 & 19.3 & 19.8 \\ \cline{2-6} 
 & QAOA-10 p4 & 31.1 & 29.9 & 13.7 & 11.1 \\ \cline{2-6} 
 & QAOA-12 p4 & 14.1 & 26.6 & 5.12 & 2.94 \\ \cline{2-6} 
 & QAOA-14 p2 & 11.5 & 14.3 & 5.93 & 4.49 \\\hline

\end{tabular}}
\end{small}
\end{center}
\end{table}

\newpage
\subsection{Impact of Number of Circuits with Partial Measurements and Selection Method}
Our default design uses a sliding window method to generate a handle of unique CPM. To understand the impact of the number of CPM and selection method, we perform an empirical study using a $12$-qubit QAOA program on IBMQ-Paris. 

\begin{figure}[htb]
\centering
    \vspace{-0.1 in}
    \includegraphics[width=\columnwidth]{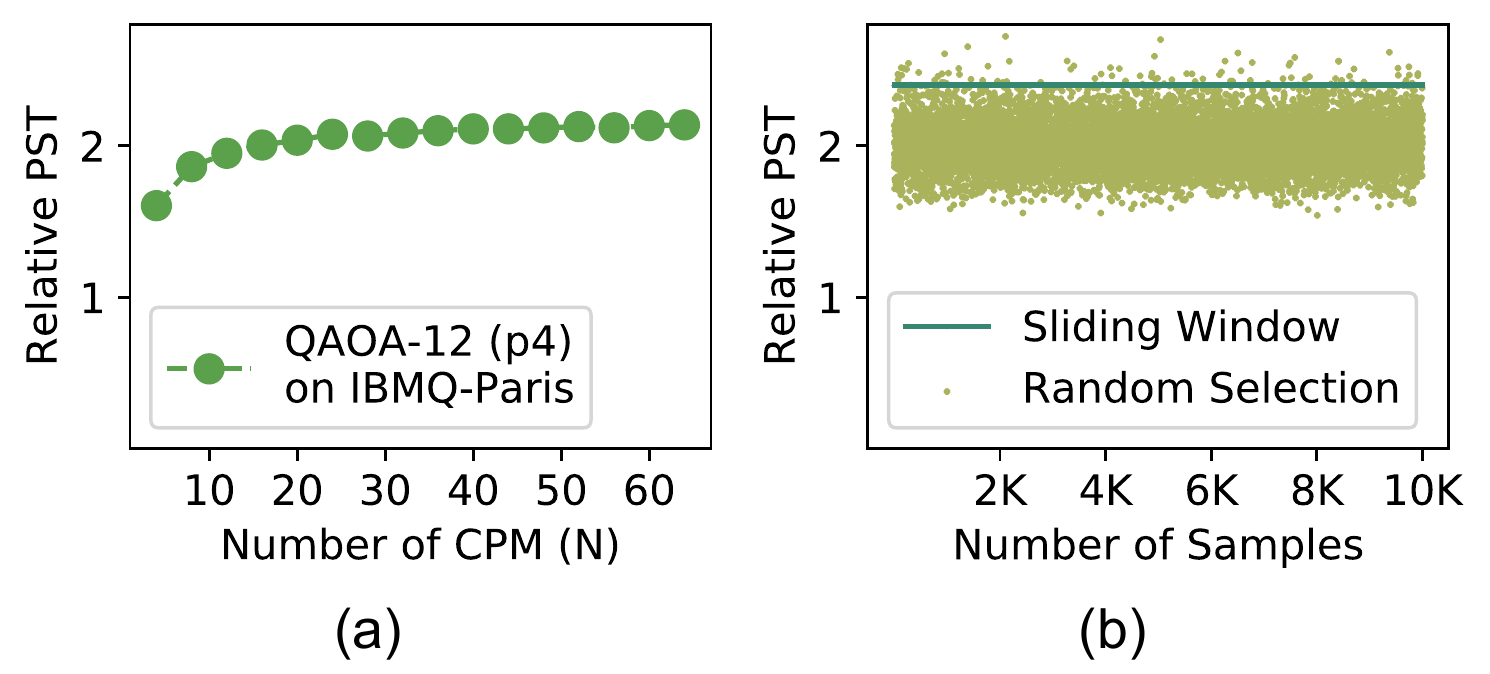}
    \vspace{-0.25 in}
    \caption{Impact of (a) Number of CPM and (b) CPM Selection Method on the Performance of JigSaw.} 
    \label{fig:sensitivitytomarginals}
    
\end{figure}

\vspace{0.05in}
\noindent \textbf{Sensitivity to Number of CPM}: The total number of possible CPM (measuring two qubits) for a Q qubit program is $^{Q}C_2$. To understand the impact of the number of CPM $(N)$ on its effectiveness, JigSaw randomly generates $\mathrm{N}$ circuits with partial measurements of subset size $2$ out of all the 66 possibilities ($^{12}C_2$ = 66) and uses these $\mathrm{N}$ local-PMFs to update the global-PMF. \ignore{Since there are $66$ CPM possible, $\mathrm{N}$ ranges between $1$ and $66$.} The process is repeated hundreds of times for each $\mathrm{N}$ and Figure~\ref{fig:sensitivitytomarginals}(a) shows the average improvement in Application PST from JigSaw as $\mathrm{N}$ is increased. 

We observe that the performance of JigSaw saturates when additional CPM do not offer incremental information. Thus, only a few and unique CPM are sufficient for JigSaw to be effective. Further, to obtain a greater number of unique CPM, JigSaw-M generates CPM of non-uniform subset sizes.

\vspace{0.05in}
\noindent \textbf{Sensitivity to CPM Selection Method}: To study the impact of CPM selection method, JigSaw randomly selects a group of CPM of subset size 2 from all the 66 possibilities while ensuring that each program qubit is measured in a CPM at least once. As there are 12 qubits in the program JigSaw selects 12 random CPM each time and the process is repeated 10,000 times. Figure~\ref{fig:sensitivitytomarginals}(b) shows the relative improvement in PST for this study, and we observe that we get similar results irrespective of the CPM. Thus, although our default design uses a sliding window method, JigSaw is equally effective even if any other technique to generate the CPM is used.

\subsection{Impact of Recompilation} 
\begin{figure}[htb]
\centering
    \includegraphics[width=\columnwidth]{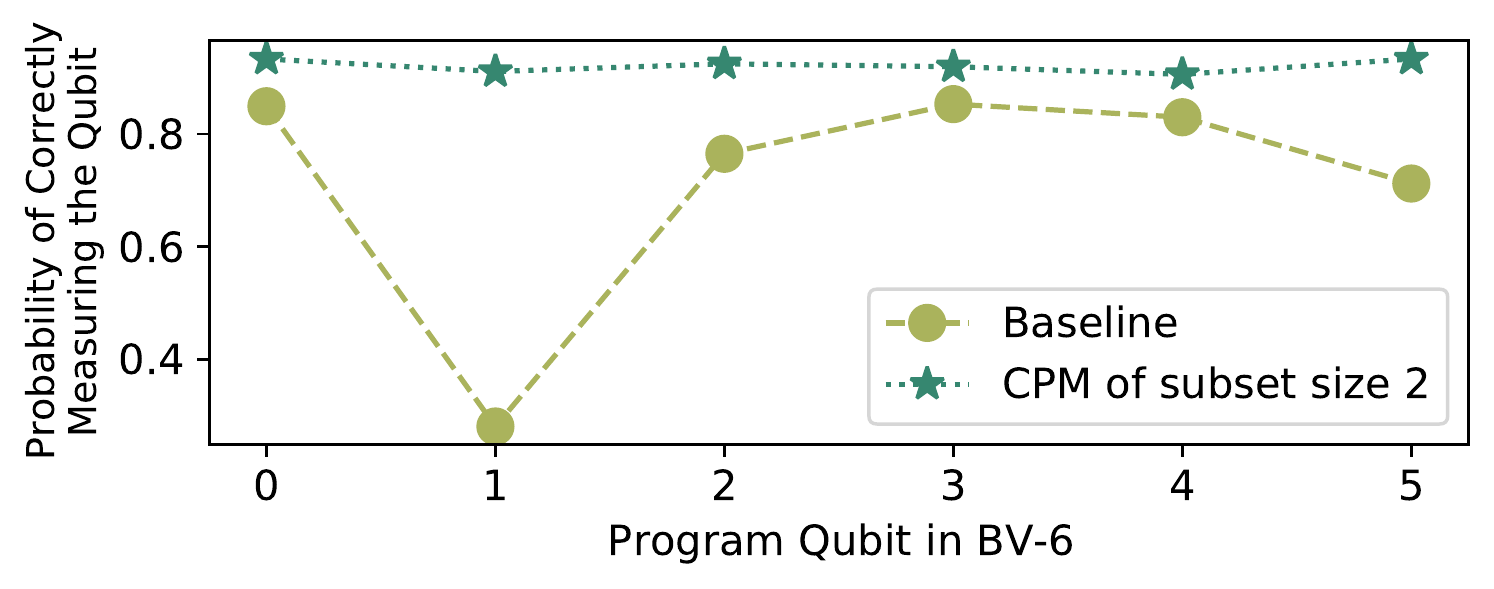}
    \caption{The probability of successfully measuring each qubit for a 6-qubit BV program on IBMQ-Toronto in the (a) baseline (b) in each CPM after recompilation.
    }
    \label{fig:impactofrecompilation}

\end{figure}

JigSaw mainly benefits from measurement subsetting and recompilation.
By recompiling each CPM, the effective measurement error-rates for CPM are close to the best-case qubits rather than close to the average-case qubits (which is the case for the global-mode and the baseline). For example, Figure~\ref{fig:impactofrecompilation} shows that the probability of correctly measuring a qubit in a CPM increases by up-to 3.25x compared to the baseline for a BV-6 benchmark on IBMQ-Toronto. Note that the probability of correctly measuring a qubit is computed from the set of outcomes where the particular qubit is correctly measured, even if the overall outcome is erroneous and does not represent the correct answer. Thus, recompiling CPM can significantly enhance the effectiveness of JigSaw.

\begin{figure}[htb]
\centering
    \includegraphics[width=1.0\columnwidth]{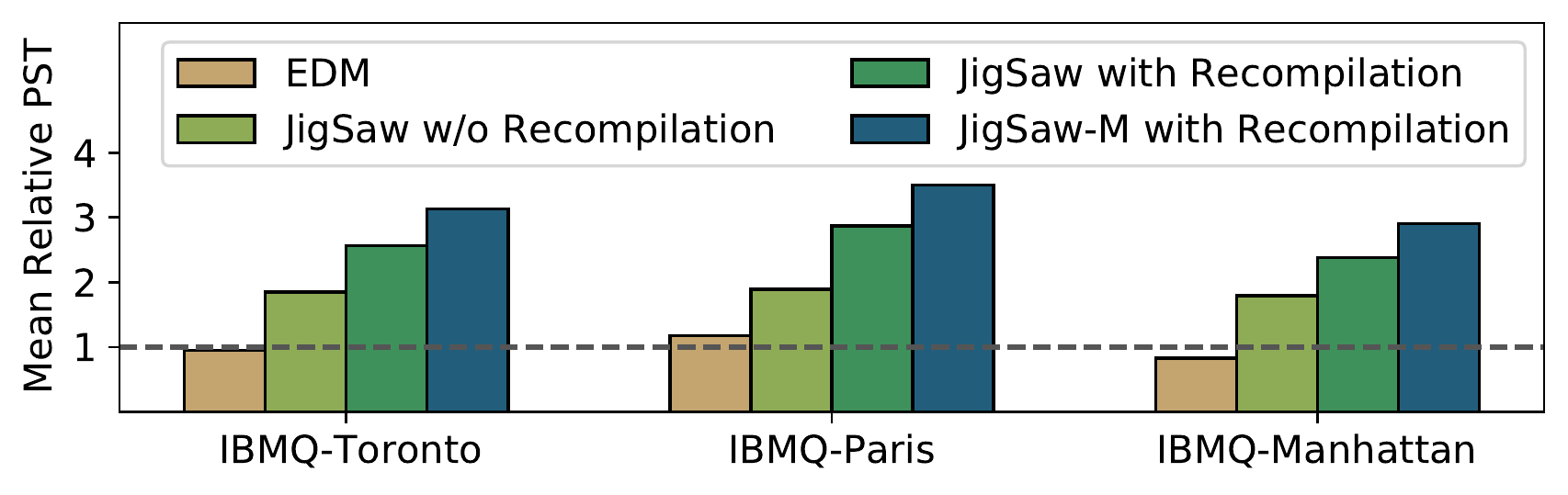}
    \caption{Comparison of Mean PST relative to the Baseline.}
    \label{fig:comparisonofallmethods}
\end{figure}

Figure~\ref{fig:comparisonofallmethods} shows the Mean PST from JigSaw without recompilation (subsetting only), JigSaw with recompilation, and JigSaw-M relative to the baseline. Without recompilation, JigSaw improves the PST by 1.92x on average and up-to 3.26x, whereas with recompilation JigSaw improves the PST by 2.91x on average and up-to 7.8x compared to the baseline. With recompilation, JigSaw-M improves the PST by 3.65x on average and by up-to 8.4x.

\vspace{-0.2in}
\section{Scalability of Reconstruction}
\label{sec:scalability}
The applicability of reconstruction algorithms is often limited by their memory and time complexity. 
For example, tensor-product based algorithms are extremely hard to scale due to their exponential complexity. Therefore, we study the scalability of our proposed Bayesian Reconstruction using an analytical model, described next. 

\subsection{Insight: Bounded by Observations}
JigSaw limits the complexity of the reconstruction algorithm by storing and updating only the non-zero entries generated in the global-mode. Although the number of possible non-zero entries in the global-PMF can scale exponentially with the program size, the actual number of entries observed is much lower and is {\em limited by the number of trials}, particularly for large programs. For example, there are $2^{100}$ possible outcomes for a $100$-qubit program, and assuming the circuit  produces a uniform distribution, it would require a minimum of $2^{100}$ or $10^{30}$ trials to observe each of these outcomes at least once, which is impractical. In practice, we may be able to execute at-most a few million trials, since the time to execute the trials still increases linearly with the number of trials. Moreover, practical quantum algorithms are designed to produce output distributions with relatively low variance and bounded possible outcomes. 
For example, Table~\ref{tab:outcomes} shows that a Graycode-18 benchmark produces only up to $18.5$K unique outcomes when executed for $512$K trials, even though $256$K (=$ 2^{18}$) outcomes are possible. We bound the complexity of JigSaw reconstruction by focusing only on the outcomes observed rather than all the possible outcomes.

\begin{table}[htb]
\begin{center}
\begin{small}
\caption{Number of Outcomes in the Global-PMF for a Graycode-18 Benchmark on IBMQ Hardware}
\setlength{\tabcolsep}{1.5mm} 
\renewcommand{\arraystretch}{1.2}
\label{tab:outcomes}
{
\begin{tabular}{ |c|c|c|c| } 
\hline
Outcomes &  IBMQ-Toronto & IBMQ-Paris & IBMQ-Manhattan \\
\hline 
\hline
Observed (Obs)  & 17.0 K & 17.3 K & 18.5 K \\
\hline   
Maximum (Max)  & 256 K & 256 K & 256 K \\
\hline 
Ratio (Obs/Max) & 6.6 \% & 6.8 \% & 7.2 \% \\
\hline 

\end{tabular}}
\end{small}
\end{center}
\end{table}

\subsection{Memory Complexity} 
JigSaw only stores non-zero entries in the global, output, local PMFs, and an intermediate PMF for each CPM. We assume $\mathrm{N}$ is the number of CPM. For simplicity, we assume the global-mode and each CPM in subset-mode uses $\mathrm{T}$ trials.\footnote{For simplicity, we assume up-to 1 million trials each for the global-mode and each CPM, which is a severely pessimistic assumption. In practice, the trials are split between a large number of CPM, so the storage and timing complexity gets reduced further.}

\begin{figure}[tb]
\centering
    \includegraphics[width=0.9\columnwidth]{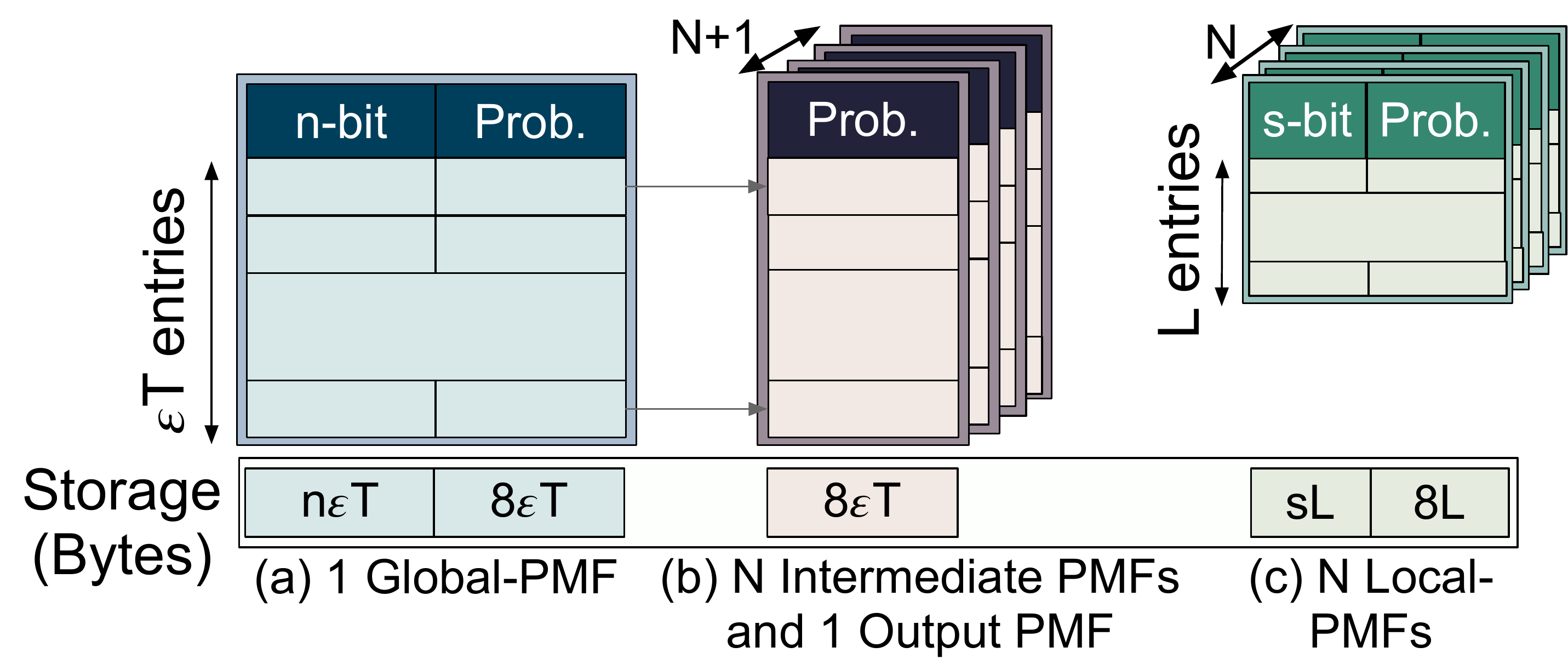}
    \vspace{-0.05 in}
    \caption{Memory required to store (a) the global, (b) N intermediate and 1 output PMF, and (c) N local-PMFs.} 
    \vspace{-0.15 in}
    \label{fig:tablesize}
\end{figure}

\vspace{0.05in}
\noindent \textbf{Global, Intermediate, and Output PMFs}: 
Each global-PMF entry comprises of an n-bit string outcome and its probability of occurrence, as shown in Figure~\ref{fig:tablesize}(a). We assume $\epsilon \mathrm{T}$ entries in the global-PMF, where $0 <  \epsilon \leq 1$.  As JigSaw only updates the probabilities of the global-PMF entries, the intermediate and output PMFs are only required to store the updated probabilities, as shown in Figure~\ref{fig:tablesize}(b). Hence, the global-PMF requires ($\mathrm{n}+8$) bytes per entry, whereas the intermediate and output PMFs each require $8$ bytes per entry. 

\begin{figure}[htb]
\centering
\vspace{-0.07 in}
    \includegraphics[width=\columnwidth]{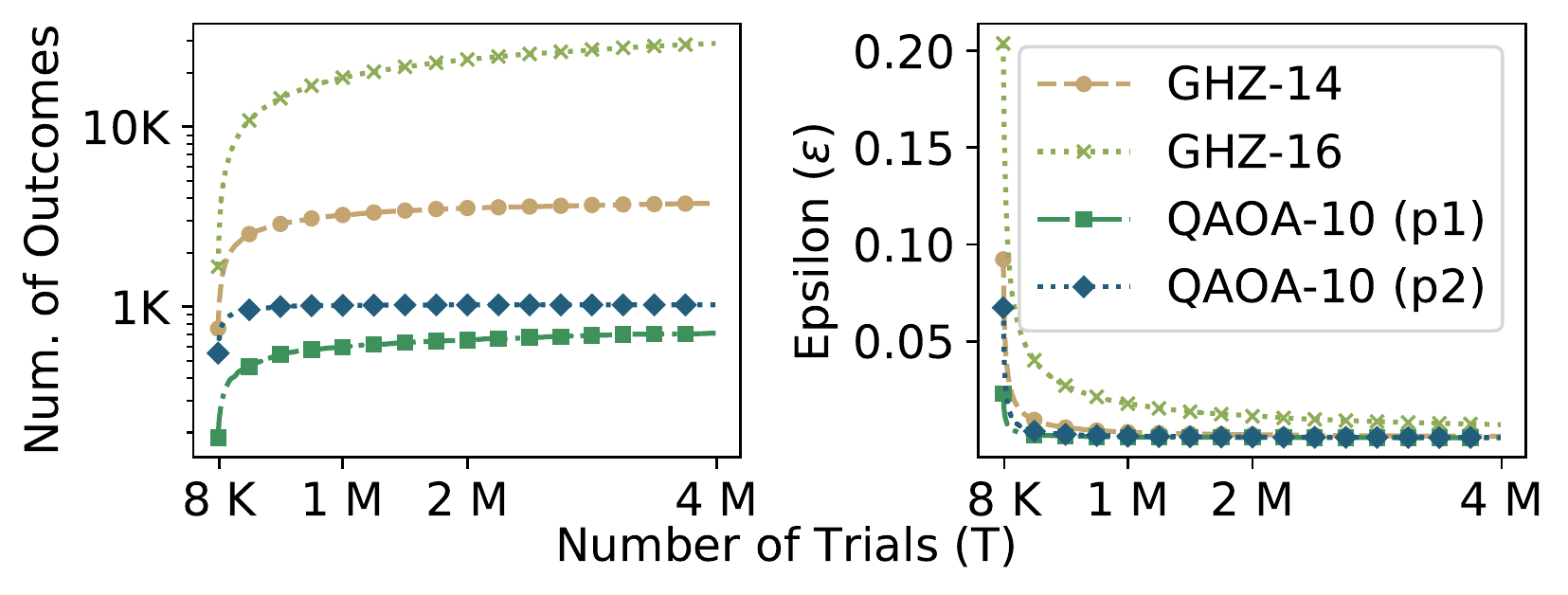}
    \vspace{-0.07 in}
    \caption{(a) Number of Global-PMF entries and (b) Epsilon ($\epsilon$) with increasing trials ($\mathrm{T}$) on IBMQ-Paris.}
    \vspace{-0.1 in}
    \label{fig:epsilon}
\end{figure}

Our experiments on IBMQ systems show that  $\epsilon\ll1$ and does not change rapidly with increasing trials. For example, Figure~\ref{fig:epsilon} shows the number of unique outcomes and $\epsilon$ when some GHZ and QAOA benchmarks are executed for $4$ $\mathrm{million}$ trials on IBMQ-Paris.

\noindent \textbf{Local-PMFs}: A local-PMF for a CPM of subset size $\mathrm{s}$ consists of $\mathrm{L} = min(2^\mathrm{s}, \delta \mathrm{T})$ entries, where $0 < \delta \leq1$, and requires $\mathrm{L}(\mathrm{s}+8)$ bytes, as shown in Figure~\ref{fig:tablesize}(c). To minimize errors on CPM, s must be small such as $2$ in the default JigSaw design. For such small $\mathrm{s}$, a local-PMF consists of all possible $2^\mathrm{s}$ entries. But for large $\mathrm{s}$, $\mathrm{L} \ll 2^\mathrm{s}$ and is denoted by $\delta \mathrm{T}$. For example, local-PMFs of size $2$ and $10$ for a GHZ-14 program on IBMQ-Toronto contain $4$ and $297$ entries respectively, even though $1024$ entries are possible for $\mathrm{s}=10$. 

JigSaw stores one Global-PMF, $\mathrm{N}$ intermediate PMFs, one output PMF and $\mathrm{N}$ local-PMFs. \ignore{and the total memory required (in bytes) to store these PMFs is given by Equation~\eqref{eq:memorycapjigsaw}.
\vspace{-0.02 in}
\begin{equation}
\label{eq:memorycapjigsaw}
{\textrm{Mem (JigSaw)} = \{\mathrm{n}+8(2+\mathrm{N})\}\epsilon \mathrm{T} + \mathrm{L}(\mathrm{s}+8)\mathrm{N}}
\end{equation}
}
JigSaw-M stores one Global-PMF, $\mathrm{N}$ intermediate PMFs, one output PMF and $\mathbb{S}\mathrm{N}$ local-PMFs where $\mathbb{S}$ subset sizes are used. Although JigSaw-M uses more CPM, since it employs hierarchical reconstruction  from the highest to the lowest subset size, only $\mathrm{N}$ intermediate PMFs are required which are reused across reconstruction rounds. Thus, the total memory capacity (in bytes) is given by Equation~\eqref{eq:memorycapjigsawm}. The memory complexity for JigSaw is obtained for $\mathbb{S}=1$ since it uses CPM of a single subset size only.
\vspace{-0.02 in}
\begin{equation}
\label{eq:memorycapjigsawm}
{\textrm{Memory} = \{\mathrm{n}+8(2+\mathrm{N})\}\epsilon \mathrm{T} + \mathrm{L}(\mathrm{s}+8)\mathbb{S}\mathrm{N}}
\end{equation}

\subsection{Time Complexity} 
JigSaw updates each Global-PMF entry for each entry in a local-PMF. Obtaining the update coefficients require $\epsilon \mathrm{T}$ operations and the update itself requires $3\epsilon \mathrm{T}$ operations per local-PMF. Assuming JigSaw uses $\mathrm{N}$ CPM, it requires $4\epsilon\mathrm{N}\mathrm{T}$ operations. Similarly, JigSaw-M requires $4\epsilon \mathbb{S}\mathrm{N}\mathrm{T}$ operations. As JigSaw only stores and updates non-zero PMF entries, which is much lower than the maximum possible, and is limited by the number of trials, the time-complexity increases linearly with the number of trials and qubits.

\subsection{Results for Scalability Analysis}
Table~\ref{tab:scalabilitynumbers} shows the memory and number of operations required for programs of different input sizes $\mathrm{n}$, values of $\epsilon$, $\delta$, and number of trials $\mathrm{T}$. To obtain the typical complexity, we use $\mathrm{T} = 1$ $\mathrm{million}$ and $\epsilon = \delta = 0.05$ (from Figure~\ref{fig:epsilon}), whereas we use $\epsilon = \delta = 1$ to obtain the upper bound. For JigSaw, we assume CPM of subset size 5 and the number of CPM ($\mathrm{N}$) to be same as the number of qubits in the program (as our default design). For JigSaw-M, we assume sizes 5,10,15, and 20. We observe that the storage and time complexity is linear with the number of trials and qubits in the program, making JigSaw applicable to programs with hundreds of qubits.  

\begin{table}[htb]
\begin{center}
\begin{small}
\ignore{\caption{
Memory (in GB) and Number of Operations (in million) required for Programs of different input sizes $\mathrm{n}$, values of $\epsilon$, $\delta$, and number of trials $\mathrm{T}$}}
\caption{Scalability Analysis of JigSaw and JigSaw-M: Memory (in GB) and Number of Operations (in million) }
\setlength{\tabcolsep}{1.5mm} 
\renewcommand{\arraystretch}{1.2}
\label{tab:scalabilitynumbers}
{
\begin{tabular}{ |c|c|c|c|c||c|c|} 
\hline
{Qubits} & \multirow{2}{*}{$\epsilon = \delta$} & Trials & \multicolumn{2}{c||}{JigSaw} & \multicolumn{2}{c|}{JigSaw-M}  \\
\cline{4-7}
(n) & & ($\mathrm{T}$)  & Mem & OPs & Mem & OPs \\
\hline
\hline
\multirow{4}{*}{100} & \multirow{2}{*}{0.05} &  32K & 0.01 & 0.66  & 0.02 & 2.64 \\
 \cline{3-7} 
& &  1024K & 0.05 & 21.0  & 0.42 & 83.9 \\
 \cline{2-7} 
& \multirow{2}{*}{1.0} &  32K & 0.03 & 13.1  & 0.20 & 52.4 \\
 \cline{3-7} 
& &  1024K & 0.96 & 419  & 3.97 & 1677 \\
 \hline 

\multirow{4}{*}{500} & \multirow{2}{*}{0.05} &  32K & 0.01 & 3.28  & 0.1 & 13.12 \\
 \cline{3-7} 
& &  1024K & 0.24 & 105  & 2.09 & 419 \\
 \cline{2-7} 
& \multirow{2}{*}{1.0} &  32K & 0.15 & 65.5  & 0.99 & 262 \\
 \cline{3-7} 
& &  1024K & 4.74 & 2097  & 19.8 & 8388 \\
 \hline

\end{tabular}}
\end{small}
\end{center}
\end{table}
\section{Related Work}
Software mitigation of NISQ hardware errors ~\cite{gokhale2019partial,gokhale2020optimized,zulehner2018efficient, noiseadaptive,nishio,shi2019optimized,micro1,tannu2018case,wilson2020just,micro3,huang2019statistical,murali2020architecting,murali2019full,NAE,shi2020resource,patel2020ureqa,patel2021qraft} is an active area of research. More recently, schemes that specifically reduce the impact of measurement errors have been proposed~\cite{matrixmeasurementmitigation,bravyi2020mitigating,kwon2020hybrid,FNM,barron2020measurement,patel2020disq}. We compare with two of these proposals, and then other works. 

\begin{figure}[htb]
\centering
    \includegraphics[width=\columnwidth]{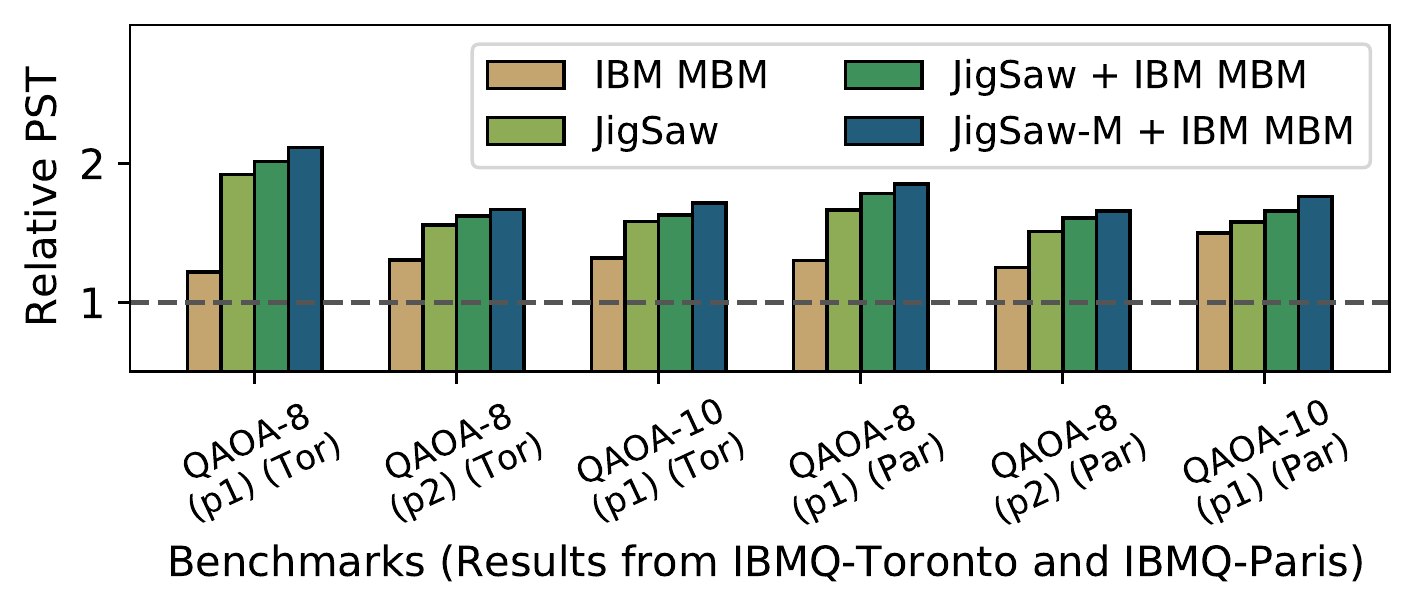}
\caption{JigSaw vs. IBM's Error Mitigation~\cite{matrixmeasurementmitigation}} 
    \label{fig:comparisonwithmem}
\end{figure}

\noindent \textbf{Matrix-Based Error Mitigation:}
IBM's matrix-based complete measurement error mitigation (MBM) post-processes the  outputs of an $\textrm{n}$-qubit program using a $2^\textrm{n}$ x $2^\textrm{n}$ inverse noise matrix prepared by calibrating $2^\textrm{n}$ basis states~\cite{matrixmeasurementmitigation}. Jigsaw can be combined with MBM for even higher fidelity than either scheme standalone, as shown in Figure~\ref{fig:comparisonwithmem}. Note that the complexity of MEM grows exponentially with the program size, whereas JigSaw needs no characterization and its post-processing step has linear complexity.

\vspace{0.05in}
\noindent \textbf{State-Based Error Mitigation:}
Few prior schemes transform a more error-prone quantum state to a less susceptible one~\cite{kwon2020hybrid,FNM} using single qubit gates, but has limited applicability on recent devices as they do not exhibit considerable bias in measuring state ``1" over state ``0". For example, the average probabilities of incorrectly measuring states “0” and “1” on IBMQ-Manhattan are 2.3\% and 3.6\%, respectively. We make similar observations on other machines too. 
\ignore{
\begin{figure}[htb]
\centering
    \includegraphics[width= \columnwidth]{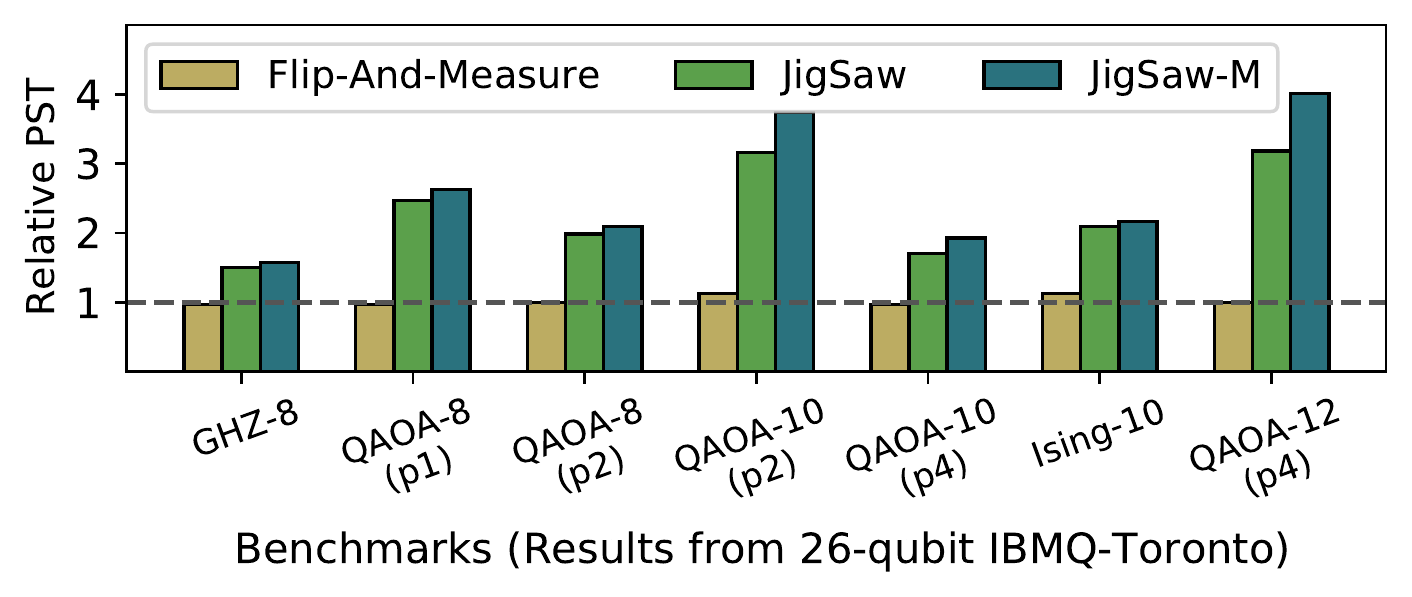}
    \caption{JigSaw vs. Flip-and-Measure policy~\cite{FNM}} 
    \label{fig:comparisonwithfnm}
\end{figure}}

\vspace{0.05in}
\noindent \textbf{Related Works in Circuit Decomposition:}
Peng et. al.~\cite{peng2019simulating} proposed circuit decomposition techniques and discussed the mathematical validity of executing a larger program on smaller quantum machines. A scalable approach for efficient circuit cutting was introduced in~\cite{tang2020cutqc}. However, these approaches use tensor products and are hard to scale. On the contrary, JigSaw utilizes circuits identical to the original program except the reduced measurements and incurs linear complexity, making it scalable and applicable to all programs.  Shehab et. al.~\cite{shehab2019noise} use partial circuits to reduce computations and improve fidelity for QAOA applications.

\section{Conclusion}
In this paper, we propose {\em JigSaw}- a design that mitigates the impact of measurement errors by executing programs in two modes: \textit{global-mode} in which the program measuring all the qubits generates a global Probability Mass Function (PMF) over all the qubits, and \textit{subset-mode} in which multiple \textit{Circuits with Partial Measurements (CPM)} produce local-PMFs over only the measured qubits. The global-PMF offers full correlation but low fidelity, whereas the local-PMFs offer higher fidelity but poor correlation. By employing Bayesian updates and the local-PMFs, JigSaw enhances the global-PMF and improves the success rate of applications on average by 2.91x and up-to 7.87x. We also propose {\em Multi-Layer JigSaw (JigSaw-M)} which uses a larger number of unique CPM of non-uniform subset sizes and employs an ordered reconstruction algorithm to enhance the global-PMF. Overall, JigSaw-M improves the success rate of applications on average by 3.65x and up-to 8.42x. 
\begin{acks}
We thank Nicolas Delfosse for helpful discussions and feedback. We also thank Sanjay Kariyappa, Ananda Samajdar, and Keval Kamdar for editorial suggestions. Poulami Das was funded by the Microsoft Research PhD Fellowship. 
This research used resources of the Oak Ridge Leadership Computing Facility at the Oak Ridge National Laboratory, which is supported by the Office of Science of the U.S. Department of Energy under Contract No. DE-AC05-00OR22725. 

\end{acks}

\appendix

\section{APPENDIX}
\subsection{Bayesian Reconstruction}
\label{sec:bayesianalgorithm}
The Bayesian Reconstruction algorithm is shown in Algorithm 1.

\setlength{\textfloatsep}{0pt}
\SetAlgoNoLine
{
\begin{algorithm}[htb]
\caption{Bayesian Reconstruction Algorithm}
\label{alg:bayesianupdate}
\SetKwInput{KwInput}{Input}                
\SetKwInput{KwOutput}{Output}              
\DontPrintSemicolon
  
  \KwInput{\textrm{\textbf{(1)} Global-PMF }$\mathrm{P} = \{\mathrm{B_x: pr_x}\}$ where $\mathrm{B_x}$ is a $\mathrm{n}$-bit outcome \textrm{\textbf{(2)} Set of \textrm{j} Marginals} $\mathrm{M = \{m_j}\}$ where $\mathrm{m_j} = [\{\mathrm{By: pr_y}\}$, $\{\mathrm{i_0 ... i_k}\}]$ for $\mathrm{k}$-bit outcome $\mathrm{B_y}$}
  \KwOutput{\textrm{PMF} $\mathrm{P_{out}} = \{\mathrm{B_x}: \mathbb{P}_x\}$, $\mathbb{P}_x \in [0,1]$}

  \SetKwFunction{FBupdate}{Bayesian\_Update}
  \SetKwFunction{FRecon}{Bayesian\_Reconstruction}
 
  \SetKwProg{Fn}{Function}{:}{\KwRet}
  \Fn{\FBupdate{$\mathrm{P}$,$\mathrm{m}$}}{
        $\mathrm{P_o} = \mathrm{P}$\;
        \For{each $(\textrm{entry }\mathrm{B_y}: \mathrm{pr_y})$ in $\mathrm{m}$:}{$\textrm{candidate} = [$ $]$\;
           \For{each $\mathrm{B_x}$ in $\mathrm{P}$:}{
                 \CommentSty{// Obtain list of outcomes in P}\;
                $\textrm{outcome} \leftarrow$ \textrm{bits in }$\mathrm{B_x}$ $\textrm{corresponding to }$ \; 
                $\textrm{qubits }$ $\{\mathrm{i_0 ... i_k}\}$\;
                candidate.append((\textrm{outcome},$\mathrm{pr_x}$))\;
                \CommentSty{// Obtain Update Coefficients}\;
                normalize candidate
            }
            \For{each \textit{outcome} in $\mathrm{candidate}$:}{
            \CommentSty{// Obtain posterior probabilities }\;
                $\mathrm{P_o}[outcome] = \frac{ \textrm{candidate}[outcome] \times  \mathrm{pr_y}}{(1-\mathrm{pr_y})}$
            
            }
            normalize $\mathrm{P_o}$ 
        }
        \KwRet $\mathrm{P_o}$\;
  }
    \SetKwProg{Fn}{Function}{:}{}
  \Fn{\FRecon{$\mathrm{P}$,$\mathrm{M}$}}{
       $\mathrm{P_{out}} = \mathrm{P}$\;
       \For{each $\mathrm{m_j}$ in $\mathrm{M}$:}{
            $\mathrm{P_{post}} = \textrm{Bayesian\_Update}(\mathrm{P,m_j}) $\;
            $\mathrm{P_{out}} = \mathrm{P_{post}} + \mathrm{P_{out}}$
       }
       normalize $\mathrm{P_{out}}$\;
        \KwRet $\mathrm{P_{out}}$\;
  }
\end{algorithm}}

\subsection{Estimate for the Number of Trials}
\label{sec:trials}
For simplicity, by default we execute the global-mode for half of the trials and the subset-mode for the remaining half. We also equally distribute the trials in the subset-mode between the CPM for both JigSaw and JigSaw-M. However, if the trials are severely limited, (1)~the distribution of trials may be fine-tuned or (2)~the subset-mode may be executed for few thousands of extra trials (global-mode corresponds to the baseline). We perform an analysis of how the trials may be allocated for each CPM. 

Let there be $N (=2^n)$ possible outcomes for a program that measures $n$ qubits. If $p$ is the probability of observing an outcome and each of the $N$ outcomes is equally likely to appear at the end of a trial, then $p= 1/N$. The probability that a given outcome has appeared at least once after $t$ trials is then given by Equation~\eqref{eq:prob1}.  
\begin{equation}
    \label{eq:prob1}
    \mathbb{P}= [1-(1-p)^t]
\end{equation}
If $t \approx \alpha N$ and $N$ is large, $\mathbb{P}$ may be approximated as Equation~\eqref{eq:prob2}.
\begin{equation}
    \label{eq:prob2}
    \mathbb{P}= 1-e^{-\alpha}
\end{equation}
Thus, in order to obtain the given outcome at least once with probability $\mathbb{P}$, the number of trials required is given by Equation~\eqref{eq:prob3}.
\begin{equation}
    \label{eq:prob3}
    t = -\ln(1-\mathbb{P})N
\end{equation}
Hence, the total number of trials required to observe every possible outcome at least once with probability $\mathbb{P}$ is given by Equation~\eqref{eq:prob4}.
\begin{equation}
    \label{eq:prob4}
    \textrm{Total number of Trials} = -\ln(1-\mathbb{P})N^2
\end{equation}

We measure only 2 qubits in each CPM in the default JigSaw design and thus, only about 150 trials are required to ensure (with 99.99\% probability) that we obtain each possible answer at-least one time. Similarly, as JigSaw-M uses CPM of different sizes, the estimated number of trials would still range within a few thousands.
\bibliographystyle{ACM-Reference-Format}
\bibliography{sample-base}


\end{document}